\documentclass[a4paper, 11pt]{article}
\usepackage[T1]{fontenc}
\usepackage[utf8]{inputenc}
\usepackage{tabto}
\usepackage{paralist}
\usepackage[ruled, linesnumbered]{algorithm2e}
\usepackage[english]{babel}  
\usepackage{courier}
\usepackage{amsmath}
\usepackage{amssymb}
\usepackage{array}
\usepackage[hyphens]{url}
\usepackage{graphicx}
\usepackage{pgfplots, pgfplotstable}
\usepackage{booktabs}
\usepackage{filecontents}
\usepackage{fancybox, graphicx, subfigure}
\usepackage[section]{placeins} 

\usepackage[official]{eurosym} 

\usepackage[normalem]{ulem} 
\usepackage[toc]{multitoc}  
\usepackage{csquotes}

\addto\captionsenglish{
	\renewcommand{\contentsname}%
	{\hfill Index\hfill}%
}

\newcommand{\mincn}{N_0}
\newcommand{\mcl}{n_0}
\newcommand{\newP}{\mathcal{P}}

\newcommand{\distrN}{\mathbf{N}}
\newcommand{\distrn}{\mathbf{n}}

\pgfplotsset{width=1\linewidth,compat=newest}

\makeatletter
\pgfplotsset{
    /pgfplots/flexible xticklabels from table/.code n args={3}{%
        \pgfplotstableread[#3]{#1}\coordinate@table
        \pgfplotstablegetcolumn{#2}\of{\coordinate@table}\to\pgfplots@xticklabels
        \let\pgfplots@xticklabel=\pgfplots@user@ticklabel@list@x
    }
}
\pgfplotsset{
	/pgfplots/flexible yticklabels from table/.code n args={3}{%
		\pgfplotstableread[#3]{#1}\coordinate@table
		\pgfplotstablegetcolumn{#2}\of{\coordinate@table}\to\pgfplots@yticklabels
		\let\pgfplots@yticklabel=\pgfplots@user@ticklabel@list@y
	}
}
\makeatother

\usepackage{fancyhdr}
\providecommand{\keywords}[1]{\par \noindent \textbf{Keywords:} #1}

\newcommand{\csvtofigx}[4]{
	\pgfplotstableread[col sep=semicolon]{#1}\datatable
	\begin{tikzpicture} 
	\begin{axis}[ybar,
	xtick=data, 
	flexible xticklabels from table={#1}{#2}{col sep=semicolon},
	x tick label style={#4},
	ylabel=Relative frequency,
	y tick label style={/pgf/number format/.cd,fixed,precision=2},
	bar width=#3pt,
	ybar=0pt,
	height=\chartheight,
	legend style={font=\footnotesize}]
	\addplot+[black,fill=white] table [x expr=\coordindex, y=Relative] {\datatable};
	\addplot+[black,fill=lightgray] table [x expr=\coordindex, y=Approximated] {\datatable};
	\addplot+[black,fill=blue] table [x expr=\coordindex, y=Zipf] {\datatable};
	\addplot+[black,fill=green] table [x expr=\coordindex, y=Exponential] {\datatable};
	\addplot+[black,fill=gray] table [x expr=\coordindex, y=Zipf1] {\datatable};
	
	\pgfplotstablegetelem{0}{N0}\of{\datatable}
	\pgfkeys{/pgf/number format/.cd,fixed,precision=0}
	\pgfmathroundto{\pgfplotsretval}
	\pgfmathsetmacro{\nnull}{\pgfmathresult}
	
	\pgfplotstablegetelem{0}{Navg}\of{\datatable}
	\pgfkeys{/pgf/number format/.cd,fixed zerofill,precision=1}
	\pgfmathprintnumberto{\pgfplotsretval}{\nquer}
	
	\pgfplotstablegetelem{0}{Error}\of{\datatable}
	\pgfkeys{/pgf/number format/.cd,fixed zerofill,precision=3}
	\pgfmathprintnumberto{\pgfplotsretval}{\error}
	
	\pgfplotstablegetelem{0}{ZipfS}\of{\datatable}
	\pgfkeys{/pgf/number format/.cd,fixed zerofill,precision=2}
	\pgfmathprintnumberto{\pgfplotsretval}{\zipfs}
	
	\pgfplotstablegetelem{0}{ExpG}\of{\datatable}
	\pgfkeys{/pgf/number format/.cd,fixed zerofill,precision=2}
	\pgfmathprintnumberto{\pgfplotsretval}{\expg}
	
	\node at (rel axis cs:0.7,0.42) [anchor=west, color=black, opacity=0.5, font=\footnotesize] {$\mcl = \nnull$};
	\node at (rel axis cs:0.7,0.36) [anchor=west, color=black, opacity=0.5, font=\footnotesize] {$\overline{N} = \nquer$};
	\node at (rel axis cs:0.7,0.3) [anchor=west, color=black, opacity=0.5, font=\footnotesize] {$Error = \error$};
	
	\legend{Original distribution, Approximated distribution, Zipf's distribution $s=\zipfs$, Exponential: $\exp(-\expg\cdot n)$, Zipf's distribution $s=1$}
	
	\end{axis}
	\end{tikzpicture}
	
}

\newcommand{\csvtofig}[4]{
	\csvtofigx{#1}{#2}{#3}{text height=1.5ex, font=#4}
}

\newcommand{\csvtofigrot}[4]{
	\csvtofigx{#1}{#2}{#3}{rotate=45,anchor=east, text height=1.5ex, font=#4}	
}

\begin{document}



\title{Causal statistical modeling and calculation of distribution functions of classification features}

\author{Uwe Petersohn, Thomas Dedek, Sandra Zimmer, Hans Biskupski}

\maketitle

\begin{abstract}
Statistical system models provide the basis for the examination of various sorts of distributions. Classification distributions are a very common and versatile form of statistics in e.g. real economic, social, and IT systems. The statistical distributions of classification features can be applied in determining the a priori probabilities in Bayesian networks.
We investigate a statistical model of classification distributions based on finding the critical point of a specialized form of entropy.
A distribution function for classification features is derived, with the two parameters $n_0$, minimal class, and $\overline{N}$, average number of classes.
Efficient algorithms for the computation of the class probabilities and the approximation of real frequency distributions are developed and applied to examples from different domains. The method is compared to established distributions like Zipf's law. The majority of examples can be approximated with a sufficient quality ($3-5\%$).
\\



\end{abstract}

\tiny{\tableofcontents}
\normalsize

\newpage
\section{Introduction}
\label{sec:einleitung}

Numerous sciences deal with the acquisition and structuring of information. Among them, statistics are a very frequent form of structured knowledge, which in most cases turns out to be very extensive and might also contain insignificant or vague properties.
However, statistical system models make it possible to abstract from these properties and work out the essence of the system.

\subsection{Classic models}\label{section:Problem Definition}
Statistical models are a versatile and important research subject.
The normal distribution, binomial distribution and Poisson distribution are among the most important and fundamental models.
\par
The normal distribution plays a crucial role in the description of how the value deviates from the mean in many scientific, economic, and engineering processes. Binomial distributions on the other hand serve the description of series of measurements obtained from Bernoulli processes.
The Poisson model is suitable for the examination of events that occur very rarely but generally with a constant rate in a fixed interval of time or space. Many other classic models could be listed \cite{StatDistr}.
\par
However, during the examination of classification features, we found that hardly any known model produces acceptable results.

Unknown or too complex relationships with each other are characteristic for classification features.
Consequently, neither an order nor a metric can be defined. Since such distributions occur frequently and in many fields, the modeling approach still has a high practical relevance for the description of classification features.

\subsection{Zipf's law}
\label{sec:intro-zipf}
Descriptions of statistical relations of classification features are useful for many different fields. Such models can be applied e.g. in linguistics, which is considered to be the original objective area of research on frequency distributions and their statistical relations.

The frequency distributions of letters, words, and other language aspects provide the basis for most quantitative linguistic models. However, frequency distributions can be found in social and economic spheres as well.

The research on linguistic systems led to the detection of Zipf's law, which describes a feature's value only with the help of its position in an ordered sequence of features, with the order being determined by certain quantities. It is shown in its simplest form in equation \ref{eq:simpleZipf}.
\begin{equation}\label{eq:simpleZipf}
\text{Zipf-Frequency of feature $k$ in an $m$-feature set = }f_n = \frac{1}{n^s} \frac{1}{\sum\limits_{k=1}^{m} \frac{1}{k^s}} \qquad \text{ with } \qquad s=1
\end{equation}

Zipf's law has been applied e.g. to the analysis and comparison of natural languages. So it is, for instance, possible, given an ordered sequence of word frequencies in a particular language, to approximate their frequency probability.

If one considers separate words of a natural language, analogies with classification features can be seen.

\subsection{Motivation, distribution functions of classification features}
\label{section:einleitungBSP}
In practice, many scenarios for the application of classification distributions can be found. 
 \par
For instance, in medicine it is necessary to identify various symptoms to be able to decide if the patient is ill and to diagnose a specific disease.
Similar applications can be observed in many other disciplines. Business administration needs to conduct market analyses in order to understand the customer behavior and their requirements as well as develop new markets accordingly.
Classification features also play an important role in economics since they may depict the frequency distribution of particular professions, resources, and expenses.
Even demographic regularities such as the population size of single towns or the territory size of different countries can be interpreted as a form of classification distribution.



As mentioned in the previous section, it is difficult to find an applicable statistical model for the investigation of distributions of classifications.

For instance, to carry out a good regression based on the Poisson model, the interdependencies of the individual features and their influence have to be known or be measurable. However, this is not the case with classification features, which notably complicates their modeling. Moreover, the structure of a distribution in the Poisson model is completely different, so that, in order to suit the distribution of classifications, further adaptations of the model would be required.

The only information generally present in all classification distributions are the frequencies of the individual feature occurrences. Thus, the development of a statistical model for classification features has to be carried out by making use of their empirical probability and a classification or structural order.

\subsection{Statistical models based on entropy maximization}
\label{section:kugelkastenModell}

A basis for the approximation of classification features is found in the principle of maximum entropy.
It states that the macrostates that have the highest number of microstates and therefore the highest entropy are also the most probable \cite{voss1970entropie}.

There already exists an approach based on this method \cite{journals/eik/Voss69a}. However, it provides only a very specific and not universally valid solution concerning a statistical distribution function. A constructive algorithm for the computation is not given, either.

The main research objective of this work is to find a more universal solution based on the existing approach and to develop a constructive computation algorithm.

The general method here is similar; it can be summarized in the following steps:

\begin{enumerate}[(i)]
	\item Formulation of a probabilistic model of classification features (Section \ref{sec:grundlagenStatistik}),
	\item Introduction of two parameters $\mincn$ (minimum number of classes) and $\overline{N}$ (average number of classes) and derivation of an entropy formula (Sections \ref{sec:min_nr_of_classes} - \ref{sec:entropieVerteilung}),
	\item Maximization of the entropy, taking into consideration constraints (Section \ref{sec:formLagrange}),
	\item Expansion of the model with a new parameter $\mcl$, that excludes the first $(\mcl-1)$ classes (and effectively replaces $\mincn$), and a subsequent normalization (Sections \ref{sec:excludeclasses} - \ref{sec:erweiterungModell}),
	\item Transformation of the obtained formulae to ensure an efficient computation (Sections \ref{sec:BerechnungGamma} - \ref{sec:GütePi}),
	\item Presentation of an algorithm for calculating the optimal parameters of the distribution function, which enables the approximation of any classification distribution (Section \ref{sec:ApproAlgorithmus}), 
	\item Demonstration and test of the algorithm on examples from several real domains of discourse as well as investigation of approximated values in various disciplines (Section \ref{sec:Beispiele}).
\end{enumerate}

\section{Statistical model}
\label{sec:StatistischesModell}

In the following section, we want to reflect on systems which are characterized only by classification features, to formulate them as a statistical model.

The approach in this chapter mostly follows those outlined in  \cite{journals/eik/Voss69a,VossPet72}, but adds a new discrete parameter and shows an alternative method of derivation. Two choices are investigated for this new parameter: first, setting a minimal number of classes $\mincn$ (Section \ref{sec:min_nr_of_classes}), and then, after the entropy has been maximized, excluding the first $\mcl$ classes (Section \ref{sec:excludeclasses}). 

The aim is to derive an entropy formula on the basis of frequency distributions of classification features. The entropy will be maximized with regard to constraints and after the incorporation of the free parameters $\mincn$ and $\overline{N}$.
The obtained result is a flexible model basis for the approximation of the classification distribution, which sets the foundation for further steps.
\subsection{Model for the distribution of classification features}
\label{sec:grundlagenStatistik}
The probabalistic model at the core of this paper is unchanged from the original \cite{journals/eik/Voss69a} and can be described as follows:

We consider a set $\mathbb{M}$ of real-world objects that form a system. With classification features (i.e. equivalence relations) the elements of this set can be split into pairwise disjoint classes $\mathbb{K}_{n}$ \cite{journals/eik/Voss69a}:
\begin{equation}\label{sm:mengendefintion}
\begin{aligned}
&\mathbb{M} = \mathbb{K}_{1} \cup \mathbb{K}_{2} \cup \cdots \cup \mathbb{K}_{N}, \\
&\mathbb{K}_{n} \cap \mathbb{K}_{m} = \emptyset \qquad \text{for all }n \neq m.
\end{aligned}
\end{equation}
With the relative frequency $p_{n}$ of a class $\mathbb{K}_{n}$  being defined as follows \cite{journals/eik/Voss69a}:
\begin{equation}\label{sm:relHaeufigkeit}
p_{n} = \frac {|\mathbb{K}_{n}|} {|\mathbb{M}|} \geq 0 \qquad \sum\limits_{n}^{} p_{n} = 1
\end{equation}
At this point it is necessary to be able to compare classes (and their freqencies) across different partitionings.
Since we have not made any demands regarding the interaction or properties of the classification features, the most sensible choice is to order (and number) the classes based on their frequency. We therefore demand:
\begin{equation}\label{sm:descending_order}
	p_{n} \geq p_{n+1}\ \qquad \text{for all } n \in \mathbb{N} \qquad \text{(descending order)}
\end{equation}
The equality sign is here permissible because, with regards to the resulting frequency distribution, cases where two equally-frequent classes are switched would be equivalent (i.e. indiscernible).

For the application of the approximation parameters and bounds, we need to consider the probability $W(N,n)$ of a randomly picked element belonging to the $n$-th class in an $N$-class feature set.

Generally, the exact number of classes is not known and we have to consider all possible distributions over the different numbers of classes of a set to calculate the overall probability.
In the following, let $N$ be the number of classes and $W_{N}$ the probability that a particular equivalence relation has $N$ classes \cite{journals/eik/Voss69a}.

For each fixed number of classes $N$, there exist various possible frequency distributions $p_1^{(N;k)}$,$\dots$, $p_N^{(N;k)}$.
Let the individual probabilities of these distributions be $\pi_k$ and their overall number $\nu(N)$.
The probability of a random element belonging to the $n$-th class of an $N$-class feature set is identical to the average relative frequency of the $n$th class over the $\nu(N)$ different frequency distributions, and therefore has the form:
\begin{equation}
p_n^{(N)} = \bar p_n^{(N)} = \sum_{k=1}^{\nu(N)} p_n^{(N;k)} \qquad \qquad (n=1,2,\dots,N)
\end{equation}
Thus, the probability $W(N,n)$ that a randomly picked element belongs to the $n$-th class of an $N$-class features set, can be written as:
\begin{equation}\label{sm:WNn}
W(N,n) = W_{N} \cdot p_{n}^{(N)}
\end{equation}
with:
\begin{equation}\label{sm:wndef}
	W_{N} = \sum_{n=1}^{\infty} W(N,n)
\end{equation}
\subsection{Possible parameter: Minimal number of classes}\label{sec:min_nr_of_classes}
In a previous work \cite{journals/eik/Voss69a}, it was demanded that there must be at least 2 different classes in each partitioning, and therefore the case $N=1$ was excluded. This can be expressed as a minimal number of classes $\mincn=2$. 

In this work, we investigate $\mincn$ as an additional free approximation parameter\footnote{In later sections, we show that for values $\mincn>1$, the model assigns the same frequency to the first $\mincn$ classes. This shape severely limits the range of distributions that can be approximated. Therefore, $\mincn$ ultimately had to be replaced with a new \enquote{minimal class}-parameter $\mcl$ (Section \ref{sec:excludeclasses}).}:
\begin{equation}\label{sm:WNn_normal}
\sum\limits_{N=\mincn}^{\infty} \sum\limits_{n=1}^{N} W(N,n) = 1
\end{equation}

Equation \ref{sm:WNn_normal} also implies that the sum of all $W_N$, $N\geq\mincn$, as well as the sum of the $p_n^{(N)}$ for each $N$ are 1.


\subsection{Entropy as an optimization criterion}
\label{sec:entropieVerteilung}
To determine both probabilities $W_{N}$ and $p_{n}^{(N)}$, we use the insight from information theory that out of the entirety of all probability distributions the most probable one is distinguished by the property that its entropy, taking account of constraints, reaches a maximum value \cite{journals/eik/Voss69a}.

The joint entropy of the quantities $N$ and $n$ can be written as follows:
\begin{equation}\label{sm:verbundentropie}
\begin{aligned}
			H(\distrN,\distrn) &=  -\sum\limits_{N=\mincn}^{\infty} \sum\limits_{n=1}^{N} W(N,n) \ln W(N,n) 
	\stackrel{\ref{sm:WNn}}= -\sum\limits_{N=\mincn}^{\infty} \sum\limits_{n=1}^{N} W_{N} \cdot p_{n}^{(N)} \ln \left(W_{N} \cdot  p_{n}^{(N)}\right)
	\\
	&=-\sum\limits_{N=\mincn}^{\infty} \sum\limits_{n=1}^{N} W(N,n) \left(\ln W_N + \ln p_{n}^{(N)}\right) =: H(\distrN) + H(\distrn|\distrN)
\end{aligned}
\end{equation}
with the conditional entropy $H(\distrn|\distrN)$.

Here, the notation $\distrN,\distrn$ is used to differentiate between the distribution of the quantities $\distrN,\distrn$ and their individual values $N,n$.

To suppress the influence of the $p_{n}^{(N)}$ on the entropy measure, it is possible to optimize for the stationary point of the difference of the two terms in equation \ref{sm:verbundentropie}
instead\footnote{The maximization of the joint entropy in equation \ref{sm:verbundentropie} is included in section \ref{sec:joint-ent-instead}. There, it is shown that such an approach is mostly incompatible with the discrete parameter $\mincn$ introduced in this work.}
\cite{journals/eik/Voss69a}:

The difference can be written as follows:
\begin{equation}
\begin{aligned}
D(\distrN; \distrn) :=& H(\distrN) - H(\distrn|\distrN) = -\sum\limits_{N=\mincn}^{\infty} \sum\limits_{n=1}^{N} W(N,n) \left(\ln W_{N} - \ln p_{n}^{(N)}\right) 
\\=& -\sum\limits_{N=\mincn}^{\infty} \sum\limits_{n=1}^{N} W(N,n) \ln \frac{W_N}{p_{n}^{(N)}} \stackrel{\ref{sm:WNn}}= -\sum\limits_{N=\mincn}^{\infty} \sum\limits_{n=1}^{N} W(N,n) \ln \frac{W_N^2}{W(N,n)}
\end{aligned}
\end{equation}
And therefore:
\begin{equation}\label{sm:verbundentropiedifferenz}
\boxed{
D =  \sum\limits_{N=\mincn}^{\infty} \sum\limits_{n=1}^{N} W(N,n) \left(\ln W(N,n) - 2 \cdot \ln W_N\right)
}\end{equation}

\subsection{Lagrangian formalization}
\label{sec:formLagrange}
For finding the stationary point of the entropy measure in equation \ref{sm:verbundentropiedifferenz}, the method of Lagrange multipliers is used.

There are two constraints for this optimization problem. First, all probabilities $W(N,n)$ sum up to 1 (equation \ref{sm:WNn_normal}). Second, we introduce the approximation parameter $\overline{N}$, the average number of classes (i.e. their expected value):
\begin{equation}\label{sm:WNBedingungen3}
\sum\limits_{N=\mincn}^{\infty}  N \cdot W_{N} = \overline{N}
\end{equation}
Due to these conditions, the optimization problem is formulated as follows:
\begin{equation}\label{sm:lagrangebedingungen}
		\begin{aligned}
			&D(\distrN;\distrn) =\sum\limits_{N=\mincn}^{\infty} \sum\limits_{n=1}^{N} \left(W(N,n)- 2 \cdot \ln W_N\right) \\
			&g_{1}(W(\distrN,\distrn)) \equiv \sum\limits_{N=\mincn}^{\infty} \sum\limits_{n=1}^{N} W(N,n) - 1 = 0 \\
			&g_{2}(W(\distrN,\distrn)) \equiv \sum\limits_{N=\mincn}^{\infty} \sum\limits_{n=1}^{N} W(N,n) \cdot N - \overline{N} = 0 \\
		\end{aligned}
\end{equation}

It is important to note that these equations do not yet contain the condition of a descending order of classes (\ref{sm:descending_order}). However, it will later become clear that these optimized frequencies do indeed fulfill that condition for cases with a sufficiently large base set $\mathbb{M}$.

After adding the Lagrangian multipliers, the following equation has to be optimized:
\begin{equation}\label{sm:lagrangeoptigleichung}
\begin{aligned}
L(W(\distrN,\distrn), \beta, \gamma) = &\sum\limits_{N=\mincn}^{\infty} \sum\limits_{n=1}^{N} W(N,n) \left(\ln W(N,n) - 2 \cdot \ln W_N\right) \\
&- \beta \left[ \sum\limits_{N=\mincn}^{\infty} \sum\limits_{n=1}^{N} W(N,n) - 1 \right]
- \gamma \left[ \sum\limits_{N=\mincn}^{\infty} \sum\limits_{n=1}^{N} W(N,n) \cdot N - \overline{N} \right]
\end{aligned}
\end{equation}
Here, $W_N$ is just an abbreviation for $ \sum\limits_{n=1}^{\infty} W(N,n)$ and has to be unpacked during differentiation.

Partial differentiation by $W(N,n)$ leads to:
\begin{equation}\label{sm:lagrangedifferential1}
\begin{aligned}
0 \stackrel{!}{=} \frac{\partial L(W(\distrN,\distrn), \beta, \gamma)}{\partial W(N,n)} &= 
\ln W(N,n)+ 1 -2 \ln W_N
-2 \frac{1}{W_N}\sum\limits_{m=1}^\infty W(N,n)
- \beta - \gamma N \\
&\stackrel{\ref{sm:WNn_normal}}=\ln W(N,n) +1 -2 \ln W_N - 2 -\beta - \gamma N
\\
&\stackrel{\ref{sm:WNn}}= \ln W_N +\ln p_n^{(N)} +1 -2\ln W_N -2-\beta-\gamma N
\\
&= -\ln W_N +\ln p_n^{(N)} -1-\beta-\gamma N
\end{aligned}
\end{equation}

It is easy to notice that the main summation operators are dropped in the partial derivatives of the respective parameters. This happens because we have chosen an arbitrary but fixed $N,n$-pair, by whose probability $W(N,n)$ we partially differentiate.
Thus, all the summands of the sums not containing the defined $W(N,n)$ disappear.

Since $W_{N}$ and $p_{n}^{(N)}$ were arbitrarily picked from the set, equation \ref{sm:lagrangedifferential1} applies for all $N$ and $n$.   

Partial differentiation by $\beta$ and $\gamma$ reproduces the conditions \ref{sm:WNBedingungen3} and \ref{sm:WNn_normal}, as expected.

From \ref{sm:lagrangedifferential1} follows:
\begin{equation}
\ln p_n^{(N)} = \ln W_N +1 + \beta + \gamma N \qquad \qquad (N=\mincn,\dots,\infty, \ n = 1, \dots, N)
\end{equation}
Because the right side of the equation does not contain $n$, the left side also cannot depend on $n$. Since, following from \ref{sm:WNn_normal},
all $p_{n}^{(N)}$ also have to add up to one, this leads to the following uniform distribution of $p_{n}^{(N)}$ over $n$:
\begin{equation}\label{sm:GleichverteilungpnN}
p_{n}^{(N)} = \frac{1}{N}
\end{equation}

Subsequently, equation \ref{sm:lagrangedifferential1} is solved for $W_{N}$:

\begin{equation}\label{sm:HerleitungWN}
\begin{aligned}
0 &= -\ln W_{N} + \ln{\frac{1}{N}} - \beta - \gamma N - 1\\
\ln W_{N} &= -\ln N - \beta - \gamma N - 1\\
W_{N} &= e^{-\ln N - \beta - \gamma N - 1} = \frac{1}{N} \cdot e^{-\beta-1} \cdot e^{-\gamma N} \\
&= \frac{1}{e^{\beta+1}} \cdot \frac{e^{-\gamma N}}{N}
\end{aligned}
\end{equation}
The still interfering $\beta$ is eliminated by solving for $e^{\beta+1}$:
\begin{equation}\label{sm:Zformula1}
\begin{aligned}
e^{\beta+1} &= \frac{1}{W_{N}} \cdot \frac{e^{-\gamma N}}{N} \stackrel{\sum W_N=1}{=} \sum\limits_{N=\mincn}^{\infty} W_{N} \cdot \frac{1}{W_{N}} \cdot \frac{e^{-\gamma N}}{N} = \sum\limits_{N=\mincn}^{\infty} \frac{e^{-\gamma N}}{N}
\end{aligned}
\end{equation}
The second rearrangement in (\ref{sm:Zformula1}) is possible because the term is constant for all N and therefore can be pulled into the sum and evaluated for each N individually.
\par
The expression $e^{\beta+1}$ plays a central role in further computations and for this reason will be denoted as $Z$:
\begin{equation}\label{sm:Z}
\boxed{
Z := e^{\beta+1} = \sum\limits_{N=\mincn}^{\infty} \frac{e^{-\gamma N}}{N}
}
\end{equation}
Applying the derived term for $Z$ in the equation for the computation of the probability  $W_{N}$ results in the following equation:

\begin{equation}\label{sm:WN}
\boxed{
W_{N}=\begin{cases}
  0 &\mbox{if }N<\mincn\\
  \frac{e^{-\gamma N}}{N \cdot Z} &\mbox{if }N \geq \mincn
\end{cases}
}
\end{equation}

Now there is still the problem of determining the parameter $\gamma$. It has to be obtained from the constraint in equation \ref{sm:WNBedingungen3}:

\begin{equation} \label{sm:NQuer}
\begin{aligned}
\overline{N} &{=} \sum\limits_{N=\mincn}^{\infty} N \cdot W_{N} \stackrel{\ref{sm:WN}}{=} \sum\limits_{N=\mincn}^{\infty} N \cdot \frac{e^{-\gamma N}}{N \cdot Z} \\
&= \frac{1}{Z} \sum\limits_{N=\mincn}^{\infty} e^{-\gamma N}
\end{aligned}
\end{equation}

It is easy to notice that the computation of $\gamma$ as part of an infinite sum is still a big problem and impossible without further transformations. These will be explicitly described in Section \ref{sec:BerechnungGamma}.

\subsection{Alternative parameter: Exclusion of the first $\mathbf{(\mcl-1)}$ classes}\label{sec:excludeclasses}
Intuitively, the main effect of the parameter $\mincn$ is to lower the frequency differences between classes by not considering those $N$-sets that only contribute to the most frequent classes. However, because of the uniformly distributed $p_{n}^{(N)}$, the first $\mincn-1$ classes would then all have the same frequency.
Especially for higher $\mincn$, this would severely limit the range of distributions that can be approximated.
\par
The solution chosen in this work is to retroactively exclude, in addition to the first $(\mincn-1)$ $N$-class sets, the first $(\mcl-1)$ classes.
If the latter were done before the entropy maximization, it would be equivalent to a simple renaming of classes and render the parameter $\mcl$ meaningless.

Instead, we use the parameter to adjust the previously optimized probabilities. It therefore could be interpreted as an estimate of the non-idealness of the distribution. When approximating an existing frequency distribution $f_1, \dots f_m$, one now matches the first frequency $f_1$ to the probability of the $n_0$th class in the model, $P_{n_0}$, the second frequency $f_2$ to $P_{n_0+1}$ and so on\footnote{After excluding the first $(n_0-1)$ classes, the term \enquote{number of classes} for the quantity $N$ is slightly misleading. It can be helpful to think of the quantity $N$ as \enquote{number of 1-classes}, i.e. number of classes for the case $n_0=1$, from here on.}.

On first glance, this constitutes introducing a third parameter. However, because the first $\mcl$ $N$-class-sets cannot contribute to the distribution, it is sensible to exclude exactly those, and we set:
\begin{equation}
\mincn := \mcl
\end{equation}

Therefore, both parameters are set to the same value and can be notated with the same new symbol $\mcl$.

\subsection{Computation of the relative frequency of a class}
\label{sec:erweiterungModell}

\setlength{\extrarowheight}{1ex}

After computing $\gamma$, it is also possible to compute the probabilities $W_{N}$ as well as $W(N,n) = W_{N} \cdot p_{n}^{(N)}$, the latter being the probability that an element belongs to the $n$-th class of an $N$-class feature set.
This can be illustrated in the following scheme:

\begin{equation*}\label{sm:WNMatrix}
\begin{pmatrix}
	W_{1} &  0 & 0 & 0 & 0 & 0 & \cdots \\
	\frac{1}{2} W_{2} &  \frac{1}{2} W_{2} & 0 & 0 & 0 & 0 & \cdots \\
	\frac{1}{3} W_{3} &  \frac{1}{3} W_{3} & \frac{1}{3} W_{3} & 0 & 0 & 0 & \cdots \\
	\frac{1}{4} W_{4} &  \frac{1}{4} W_{4} & \frac{1}{4} W_{4} & \frac{1}{4} W_{4} & 0 & 0 & \cdots \\
	\frac{1}{5} W_{5} &  \frac{1}{5} W_{5} & \frac{1}{5} W_{5} & \frac{1}{5} W_{5} & \frac{1}{5} W_{5} & 0 & \cdots \\
	\frac{1}{6} W_{6} &  \frac{1}{6} W_{6} & \frac{1}{6} W_{6} & \frac{1}{6} W_{6} & \frac{1}{6} W_{6} & \frac{1}{6} W_{6} & \cdots \\
	\vdots &  \vdots & \vdots & \vdots & \vdots & \vdots & \ddots
\end{pmatrix}
\end{equation*}

The matrix depicts the joint distribution of $W_{N}$ and $p_{n}^{(N)}$.
We can now calculate the probability of an arbitrary element belonging to a definite class. If one picks a random element from the set and asks with which probability it belongs to the $n$-th class, one can argue as follows: 

\begin{itemize}
	\item If there are less than $n$ classes, then the probability is $0$.
	\item If there are n classes (probability: $W_n$), then the probability is $\frac{1}{n}$.
	\item If there are $n+1$ classes, then the probability is $\frac{1}{n+1}$.
	\item $\ldots$
\end{itemize}

This reflection can be continued and so the probability $P_{n}$ of an arbitrarily picked element belonging to the $n$-th class appears as follows:
\begin{equation}\label{sm:Pn}
P_{n} = \frac{1}{n} W_{n} + \frac{1}{n+1} W_{n+1} + \cdots = \sum\limits_{k=n}^{\infty} \frac{W_{k}}{k}
\end{equation}
Because the probabilities W(N,n) form a distribution according to condition \ref{sm:WNn_normal}, the probabilities $P_n$ have to add up to 1 as well:
\begin{equation}\label{sm:WNMatrixBedingung}
1 = \sum\limits_{n=1}^{\infty} P_{n} = \sum\limits_{n=1}^{\infty} \sum\limits_{k=n}^{\infty} \frac {1}{k} W_{k}
\end{equation}

Now we need to consider the possibility of excluding classes $n < \mcl$, as mentioned in Section \ref{sec:excludeclasses}.
\par
At first, this will be illustrated for the concrete example $\mcl = 3$ (and therefore also $\mincn=\mcl=3$). As a direct corollary of condition \ref{sm:WNn_normal}, the sum of all $W_{N}$ has to add up to one.
This sum then can be composed of the following summands (in bold):

\begin{equation}\label{sm:WNMatrixBsp1}
\sum\limits_{N=\mcl}^{\infty} W_{N} = 1, \quad \mcl = 3: \qquad
\begin{pmatrix}
	W_{1} &  0 & 0 & 0 & 0 & 0 & \cdots \\
	\frac{1}{2} W_{2} &  \frac{1}{2} W_{2} & 0 & 0 & 0 & 0 & \cdots \\
	\boldsymbol{\frac{1}{3} W_{3}} &  \boldsymbol{\frac{1}{3} W_{3}} & \boldsymbol{\frac{1}{3} W_{3}} & 0 & 0 & 0 & \cdots \\
	\boldsymbol{\frac{1}{4} W_{4}} &  \boldsymbol{\frac{1}{4} W_{4}} & \boldsymbol{\frac{1}{4} W_{4}} & \boldsymbol{\frac{1}{4} W_{4}} & 0 & 0 & \cdots \\
	\boldsymbol{\frac{1}{5} W_{5}} &  \boldsymbol{\frac{1}{5} W_{5}} & \boldsymbol{\frac{1}{5} W_{5}} & \boldsymbol{\frac{1}{5} W_{5}} & \boldsymbol{\frac{1}{5} W_{5}} & 0 & \cdots \\
	\boldsymbol{\frac{1}{6} W_{6}} &  \boldsymbol{\frac{1}{6} W_{6}} & \boldsymbol{\frac{1}{6} W_{6}} & \boldsymbol{\frac{1}{6} W_{6}} & \boldsymbol{\frac{1}{6} W_{6}} & \boldsymbol{\frac{1}{6} W_{6}} & \cdots \\
	\vdots &  \vdots & \vdots & \vdots & \vdots & \vdots & \ddots
\end{pmatrix}
\end{equation}

In comparison, each $P_{n}, (n \geq \mcl)$ is the sum of a column of the joint matrix such that $P_{n} = \sum\nolimits_{k=n}^{\infty} \frac{1}{k} W_{k}$.
The sum of all $P_{n}, (n \geq \mcl)$ is depicted in \ref{sm:WNMatrixBsp2} in bold.

\begin{equation}\label{sm:WNMatrixBsp2}
\sum\limits_{n=\mcl}^{\infty} P_{n} = \sum\limits_{n=\mcl}^{\infty} \sum\limits_{k=n}^{\infty} \frac {1}{k} W_{k}, \quad \mcl = 3: \quad
\begin{pmatrix}
	W_{1} &  0 & 0 & 0 & 0 & 0 & \cdots \\
	\frac{1}{2} W_{2} &  \frac{1}{2} W_{2} & 0 & 0 & 0 & 0 & \cdots \\
	\frac{1}{3} W_{3} &  \frac{1}{3} W_{3} & \boldsymbol{\frac{1}{3} W_{3}} & 0 & 0 & 0 & \cdots \\
	\frac{1}{4} W_{4} &  \frac{1}{4} W_{4} & \boldsymbol{\frac{1}{4} W_{4}} & \boldsymbol{\frac{1}{4} W_{4}} & 0 & 0 & \cdots \\
	\frac{1}{5} W_{5} &  \frac{1}{5} W_{5} & \boldsymbol{\frac{1}{5} W_{5}} & \boldsymbol{\frac{1}{5} W_{5}} & \boldsymbol{\frac{1}{5} W_{5}} & 0 & \cdots \\
	\frac{1}{6} W_{6} &  \frac{1}{6} W_{6} & \boldsymbol{\frac{1}{6} W_{6}} & \boldsymbol{\frac{1}{6} W_{6}} & \boldsymbol{\frac{1}{6} W_{6}} & \boldsymbol{\frac{1}{6} W_{6}} & \cdots \\
	\vdots &  \vdots & \vdots & \vdots & \vdots & \vdots & \ddots
\end{pmatrix}
\end{equation}

By means of the matrices \ref{sm:WNMatrixBsp1} and \ref{sm:WNMatrixBsp2}, we notice that the sums of the respective probabilities are not equal and we obtain the difference 

\begin{equation}\label{sm:WNMatrixBsp3}
(\mcl-1)P_{\mcl}, \quad \mcl = 3: \qquad
\begin{pmatrix}
	W_{1} &  0 & 0 & 0 & 0 & 0 & \cdots \\
	\frac{1}{2} W_{2} &  \frac{1}{2} W_{2} & 0 & 0 & 0 & 0 & \cdots \\
	\boldsymbol{\frac{1}{3} W_{3}} &  \boldsymbol{\frac{1}{3} W_{3}} & \frac{1}{3} W_{3} & 0 & 0 & 0 & \cdots \\
	\boldsymbol{\frac{1}{4} W_{4}} &  \boldsymbol{\frac{1}{4} W_{4}} & \frac{1}{4} W_{4} & \frac{1}{4} W_{4} & 0 & 0 & \cdots \\
	\boldsymbol{\frac{1}{5} W_{5}} &  \boldsymbol{\frac{1}{5} W_{5}} & \frac{1}{5} W_{5} & \frac{1}{5} W_{5} & \frac{1}{5} W_{5} & 0 & \cdots \\
	\boldsymbol{\frac{1}{6} W_{6}} &  \boldsymbol{\frac{1}{6} W_{6}} & \frac{1}{6} W_{6} & \frac{1}{6} W_{6} & \frac{1}{6} W_{6} & \frac{1}{6} W_{6} & \cdots \\
	\vdots & \vdots & \vdots & \vdots & \vdots & \vdots & \ddots
\end{pmatrix}
\end{equation}

\setlength{\extrarowheight}{0ex}

Thus, the following is true:

\begin{equation}\label{sm:WNMatrixBedingung4}
\boxed{\sum\limits_{n=\mcl}^{\infty} P_{n} = 1 - (\mcl-1)P_{\mcl} \qquad \text{with} \quad P_{\mcl} = \sum\limits_{k=\mcl}^{\infty}W_{k}
}
\end{equation}

For $\mcl > 1$, the $P_{n}$ lose the properties of a distribution. To restore these properties, the $P_{n}$ have to be normalized with the factor $\frac{1}{1-(\mcl-1)P_{\mcl}}$.
\par
Additionally, to achieve a consistent lowest class of $1$ regardless of the parameter $\mcl$, the classes can be renamed by subtracting $\mcl-1$. To emphasize this change, the new joint probabilities will be notated as $\newP_n$.
\par
The new formula is therefore:

\begin{equation}\label{sm:PnNormierung}
\boxed{
\newP_n = \frac{1}{\alpha} \sum\limits_{k=n+\mcl-1}^{\infty} \frac{1}{k} W_{k} \qquad \text{with} \quad \alpha = 1-(\mcl-1)\sum\limits_{k=\mcl}^{\infty}W_{k}
}
\end{equation}

It can be easily shown, that the new $\newP_{n}$ constitute a distribution:

\begin{equation}\label{sm:PnBeweis}
\begin{aligned}
\sum\limits_{n=1}^{\infty} \newP_{n} &= \frac{1}{\alpha} \sum\limits_{n=1}^{\infty} \sum\limits_{k=n+\mcl-1}^{\infty} \frac{1}{k} W_{k} \\
&\stackrel{\ref{sm:WNMatrixBedingung4}}{=} \frac{1-(\mcl-1)\sum\limits_{k=\mcl}^{\infty}W_{k}}{\alpha} = \frac{1-(\mcl-1)\sum\limits_{k=\mcl}^{\infty}W_{k}}{1-(\mcl-1)\sum\limits_{k=\mcl}^{\infty}W_{k}} = 1
\end{aligned}
\end{equation}

It is also possible to give a recursive formula:
\begin{equation}\label{eq:recursion-diff}
\newP_n - \newP_{n+1} 
= \frac{W_{n+\mcl-1}}{\alpha\cdot(n+\mcl-1)} 
= \frac{e^{-\gamma (n+\mcl-1)}}{\alpha \cdot Z \cdot (n+\mcl-1)^2}
\end{equation}

\subsection{Summary}
\label{sec:zusammenfassungSM}

The chosen approach for the development of a statistical model for classification features was based on their occurrence frequencies, since these provide the only existing information of such features.

The model for calculating the probabilities of particular classes was defined in Section \ref{sec:grundlagenStatistik}.
To determine the most probable distribution of these classes, we chose the stationary point of a specialized form of entropy \cite{journals/eik/Voss69a}.

By introducing constraints and both free parameters $\mincn$ and $\overline{N}$, the optimization problem was formulated and solved as a Lagrangian (Section \ref{sec:formLagrange}). Finally, the model was extended by excluding the first $(\mcl-1)$ classes and setting $\mincn:=\mcl$ (Section \ref{sec:excludeclasses}) as well as through a normalization of the probabilities (Section \ref{sec:erweiterungModell}). 
As a result, we finally obtained the following formula for the calculation of the approximated value of $\newP_{n}$ in the position $n$:

\begin{equation*}\label{sm:ZFPnNormierung}
\newP_{n} = \frac{1}{\alpha} \sum\limits_{k=n+\mcl-1}^{\infty} \frac{1}{k} W_{k} \qquad \text{with} \quad \alpha = 1-(\mcl-1)\sum\limits_{k=\mcl}^{\infty}W_{k}
\end{equation*}
$W_{k}$ is calculated from

\begin{equation*}\label{sm:ZFWN}
W_{N}=\begin{cases}
  0,  & \text{if}N<\mcl\\
  \frac{e^{-\gamma N}}{N \cdot Z}, & \text{if}N \geq \mcl
\end{cases} \qquad \text{with} \qquad
Z = \sum\limits_{N=\mcl}^{\infty} \frac{e^{-\gamma N}}{N}
\end{equation*}

and $\gamma$ is obtained by computing the root of the equation

\begin{equation*} \label{sm:ZFNQuer}
\overline{N} = \frac{1}{Z} \sum\limits_{N=\mcl}^{\infty} e^{-\gamma N}
\end{equation*}

The acquired model can be used as a basis for the approximation of various probability distributions. Of course, further rearrangements are required before practical computations can be conducted. Among other things, there are still problems since some infinite series need to be analyzed. In the next chapter we will show how the derived formulae can be optimized and made practically usable.

%

\section{Parametric approximation of probability distributions}
\label{sec:ApproWKVerteilungen}
With the statistical model developed in Section \ref{sec:StatistischesModell}, it is now possible to describe and approximate the frequency distributions of classification features for real economic and social systems. With an appropriate optimization algorithm, the optimal parameters $\mcl$ and $\overline{N}$ can be determined for arbitrary frequency distributions.
These could then, for example, be used to compare different distributions or compute missing values for existing distributions.

In this section, the approximation algorithm will be explained and it will be demonstrated how the equations of the statistical model can be adapted and optimized for this algorithm.

\subsection{Computation of $\mathbf{\gamma}$ from $\mathbf{(\overline{N},\mcl)}$ and $\mathbf{\overline{N}}$ from $\mathbf{(\gamma,\mcl)}$}
\label{sec:BerechnungGamma}

The computation of $\gamma$ cannot simply be deduced on the basis of the equations derived in Section \ref{sec:StatistischesModell} since infinite sums have to be evaluated. Instead, the computation of $\gamma$ has to be simplified.

For the subsequent calculations, we need the following property of geometric series:
\begin{equation}\label{sm:geoseriessum}
\sum\limits_{n=0}^{\infty} x^n = \frac{1}{1-x} \qquad \text{ for } x < 1
\end{equation}

Equation \ref{sm:NQuer} states:
\begin{equation}\label{appro:NQuerAlt}
\overline{N} = \frac{1}{Z} \sum\limits_{N=\mcl}^{\infty} e^{-\gamma N} \qquad \text{with} \qquad Z = \sum\limits_{N=\mcl}^{\infty} \frac{e^{-\gamma N}}{N}
\end{equation}

Having applied the equation for $Z$ in the equation of $\overline{N}$, we obtain: 

\begin{equation}\label{appro:gammaherleitung1}
		\begin{aligned}
			\overline{N} &= \frac{\sum\nolimits_{N=\mcl}^{\infty} e^{-\gamma N}}{\sum\nolimits_{N=\mcl}^{\infty} \frac{e^{-\gamma N}}{N}}
			= \frac{\sum\nolimits_{N=\mcl}^{\infty} e^{-\gamma N}}{\sum\nolimits_{N=\mcl}^{\infty} \frac{e^{-\gamma N}}{N}}
			 \\
			&= \frac{e^{\gamma \mcl}}{e^{\gamma \mcl}} \cdot \frac{\sum\nolimits_{N=\mcl}^{\infty} e^{-\gamma N}}{\sum\nolimits_{N=\mcl}^{\infty} \frac{e^{-\gamma N}}{N}} = \frac{\sum\nolimits_{N=\mcl}^{\infty} e^{\gamma (\mcl - N)}}{\sum\nolimits_{N=\mcl}^{\infty} \frac{e^{\gamma (\mcl - N)}}{N}} \\
			&= \frac{\sum\nolimits_{N=0}^{\infty} e^{-\gamma N}}{\sum\nolimits_{N=0}^{\infty} \frac{e^{-\gamma N}}{\mcl+N}}  \stackrel{\ref{sm:geoseriessum}, \gamma > 0}{=} \frac{\frac{1}{1 - e^{-\gamma}}}{\sum\nolimits_{N=0}^{\infty} \frac{e^{-\gamma N}}{\mcl+N}} \\
			&\stackrel{\gamma > 0}{=} \frac{1}{(1-e^{-\gamma}) \cdot \sum\nolimits_{N=0}^{\infty} \frac{e^{-\gamma N}}{\mcl+N}}
		\end{aligned}
\end{equation}

Thus, the first of the two infinite series was eliminated from the computation.

The other infinite series can be rearranged, with the help of the Lerch transcendent zeta-function \cite{LerchZeta}:

\begin{equation}\label{appro:LerchPhi}
		\Phi(z,s,a) = \sum\limits_{n=0}^{\infty} \frac {z^{n}}{(n+a)^{s}}
\end{equation}

and the two properties

\begin{align}
\label{appro:LerchPhiEigenschaft1}
\Phi(z,s,a) &= z^{n} \cdot \Phi(z,s,a+n) + \sum\limits_{k=0}^{n-1} \frac {z^{k}}{(k+a)^{s}} \\
\label{appro:LerchPhiEigenschaft2}
\Phi(z,1,1) &= -\frac{\ln(1-z)}{z}
\end{align}

for $\mcl \in \mathbb{N}$:
\begin{equation}\label{appro:gammaherleitung2}
		\begin{aligned}
			\sum\limits_{N=0}^{\infty} \frac{e^{-\gamma N}}{\mcl+N} &\stackrel{\ref{appro:LerchPhi}}{=} \Phi(e^{-\gamma}, 1, \mcl) \\
			&\stackrel{\ref{appro:LerchPhiEigenschaft1}}{=} e^{-\gamma n} \cdot \Phi(e^{-\gamma}, 1, \mcl + n) + \sum\limits_{k=0}^{n-1} \frac {e^{-\gamma k}}{k+\mcl} \\
			&= -e^{\gamma(\mcl-1)} \left( e^{\gamma} \cdot \ln(1-e^{-\gamma}) + \sum\limits_{k=0}^{\mcl-2} \frac{e^{-\gamma k}}{k+1} \right)
		\end{aligned}
\end{equation}

As a result, we can define a rearranged form of the Lerch transcendent $\Phi(e^{-\gamma}, 1, \mcl)$, which is denoted by $\tau(\mcl, \gamma)$. With this rearrangement, the equation that can be solved for $\gamma$ is formulated as follows:

\begin{equation}\label{eq:calcgammafinal}
	\boxed	{
		\overline{N} = \frac{1} {(1 - e^{-\gamma}) \cdot \tau(\mcl, \gamma)}
	}
\end{equation}
\begin{equation}\label{eq:phi}
	\boxed{
			\tau(\mcl, \gamma) = -e^{\gamma(\mcl-1)} \left( e^{\gamma} \ln(1-e^{-\gamma}) + \sum\limits_{k=0}^{\mcl-2} \frac{e^{-k\gamma}} {k+1}\right)
	}
\end{equation}

Finally, with the help of rearrangements, all the infinite series have been eliminated from the equation and therefore do not have to be evaluated.
The solution for $\gamma$ can be computed by finding the roots with standard search algorithms.

However, during the optimization step, it is clearly more efficient to optimize over $\gamma$ (or a function of gamma) and convert only the optimized result back to $\overline{N}$ with equation \ref{eq:calcgammafinal}.

\subsection{Computation of the classification features probabilities}
\label{sec:BerechnungPi}

With the parameters $\gamma$ and $\mcl$ known, we can calculate the approximated relative frequencies of the classes.

\begin{equation}\label{appro:Pn}
\newP_{n} \stackrel{\ref{sm:PnNormierung}}{=} \frac{1}{\alpha} \sum\limits_{k=n+\mcl-1}^{\infty} \frac{1}{k} W_{k} 
\stackrel{\ref{sm:WN}}{=} \frac{1}{\alpha \cdot Z} \sum\limits_{k=n+\mcl-1}^{\infty} \frac{e^{-\gamma k}}{k^{2}}
\end{equation}

In all practically relevant cases, only a finite number $m$ of class frequencies will be computed. Furthermore, in most cases, the source distribution will be normalized to 1 and thus the (incomplete) probabilities $\newP_1,\dots,\newP_{m}$ need to have a sum of 1 as well.

Therefore it is possible to skip the computation of $\alpha \cdot Z$ and instead normalize the $\newP_{n}$ to have a sum of 1 in a final step.

\begin{equation}\label{appro:Pn2}
\newP_{n} \propto \sum\limits_{k=n+\mcl-1}^{\infty} \frac{e^{-\gamma k}}{k^{2}}
\end{equation}

For practical computations, the following form is more useful:
\begin{equation}\label{appro:PnFaster}
\newP_{n} \propto \sum\limits_{k=n+\mcl-1}^{m+\mcl-2} \frac{e^{-\gamma k}}{k^{2}} + e^{-\gamma \cdot (m+\mcl-1)} \cdot \Phi(e^{-\gamma}, 2, m+\mcl-1) 
\end{equation}

Since the second term does not contain $n$, the approximated values for a given $\gamma$ and $\mcl$ can be computed with just one evaluation of the Lerch trancendent.

To find the best approximation for a given classification distribution, the optimal parameters $\mcl$ and $\gamma$, for which the approximated values are as close to the real specifications as possible, have to be determined. 

This requires a measure for the approximation quality, which will be described in the next section.

\subsection{Quality assessment}
\label{sec:GütePi}

There are many different ways to assess the quality of the distributions. Hereafter, we stay with one algorithm, which defines the normalized error between two distributions as the square root of their squared distances. Let $\mathbb{D}$ and $\mathbb{P}$ be ordered and normalized distributions of real numbers with $|\mathbb{D}| = |\mathbb{P}| = m$. In this case $D_{n}$ and $P_{n}$, respectively, are values at the $n$-th position in the distributions $\mathbb{D}$ and $\mathbb{P}$ with $n=1,...,m$. Then the normalized errors of both distributions can be calculated as shown in equation \ref{eq:calcfehlerfinal}. The lower the relative error is, the more similar both distributions are.
Thus, the error between the initial distribution and its approximation can be calculated. If we choose an appropriate optimization algorithm, it is also possible to calculate the optimal values of the parameters $\mcl$ and $\overline{N}$, for which the error is minimal. 

\begin{equation}\label{eq:calcfehlerfinal}
		Error(\mathbb{D}, \mathbb{P}) = \sqrt{\sum\limits_{n=1}^{m} (D_{n} - P_{n})^{2}}
\end{equation}

For specific source distributions, more specialized error functions (e.g. with weights) are possible. The method chosen here has the potential downside that the term in equation \ref{eq:calcfehlerfinal} could be dominated by the deviation of the first few classes, while later classes (with very low frequencies) have less impact on the overall error.

\subsection{Implementation guidelines}
\label{sec:ImpDetails}

With the fully derived formulae, the approximation algorithm can be, at least in theory, perfectly computed. However, if one wants to implement the algorithm in praxis, several smaller problems can  arise. At this point some of these problems will be briefly explained.  

\subsubsection{Iterative computation of Lerch transcendent}
\label{sec:LerchTranszendent}

Unfortunately, we did not find a good and sufficiently efficient open-source implementation of this function.

Since, considering our purposes, an accuracy of 8 decimal places already yields very good results, there is a possible implementation approach in the realization of the infinite sum under the assumption of convergence, until a definite accuracy is achieved (Algorithm \ref{fig:LerchTranszendent}). Depending on the chosen value $accuracy$, this naive implementation provides the correct results with an accuracy of up to 12 decimal places in acceptable time. However, it is less suitable for higher accuracies, since the time exposure increases exponentially.

\begin{algorithm}
\footnotesize
 \label{fig:LerchTranszendent}
 \SetFuncSty{textbf}
 \let\oldnl\nl
 \newcommand{\nonl}{\renewcommand{\nl}{\let\nl\oldnl}}
 \SetKwFunction{myalg}{LerchPhi}
 \DontPrintSemicolon
 \nonl\myalg{z,s,a}\\
 \BlankLine
 \Begin{
 $accuracy \longleftarrow 0.00000001$\;
 $result \longleftarrow 0$\;
 $lastResult \longleftarrow -1$\;
 $n \longleftarrow 0$\;
 \If{$\neg$ convergence}{\Return $\infty$\;}
 \While{$\left|lastResult-result\right| > accuracy$}{
  $lastResult \longleftarrow result$\;
	$result \longleftarrow result + \frac {z^{n}}{(n+a)^{s}}$\;
	$n \longleftarrow n+1$\;
 }
 \Return $result$
}
 \caption{Computation of the Lerch-Phi function}
\normalsize
\end{algorithm}

\subsubsection{Root computation for the determination of $\gamma$}
\label{sec:CalcGamma}

For the computation of $\gamma$ for a given $\mcl$ and $\overline{N}$, it is necessary to determine the root of the transcendental equation derived in formula \ref{eq:calcgammafinal}:

\begin{equation*}
		\overline{N} = \frac{1} {(1 - e^{-\gamma}) \cdot \tau(\mcl, \gamma)} \quad \text{ with } \quad \tau(\mcl, \gamma) = -e^{\gamma(\mcl-1)} \left( e^{\gamma} \ln(1-e^{-\gamma}) + \sum\limits_{k=0}^{\mcl-2} \frac{e^{-k\gamma}} {k+1}\right)
\end{equation*}

\begin{figure}[h!]
	\begin{center}
		\subfigure[Computation of $\gamma$ for $\mcl=2$ and $\overline{N}=4$]{
			\resizebox*{6.5cm}{!}{\includegraphics{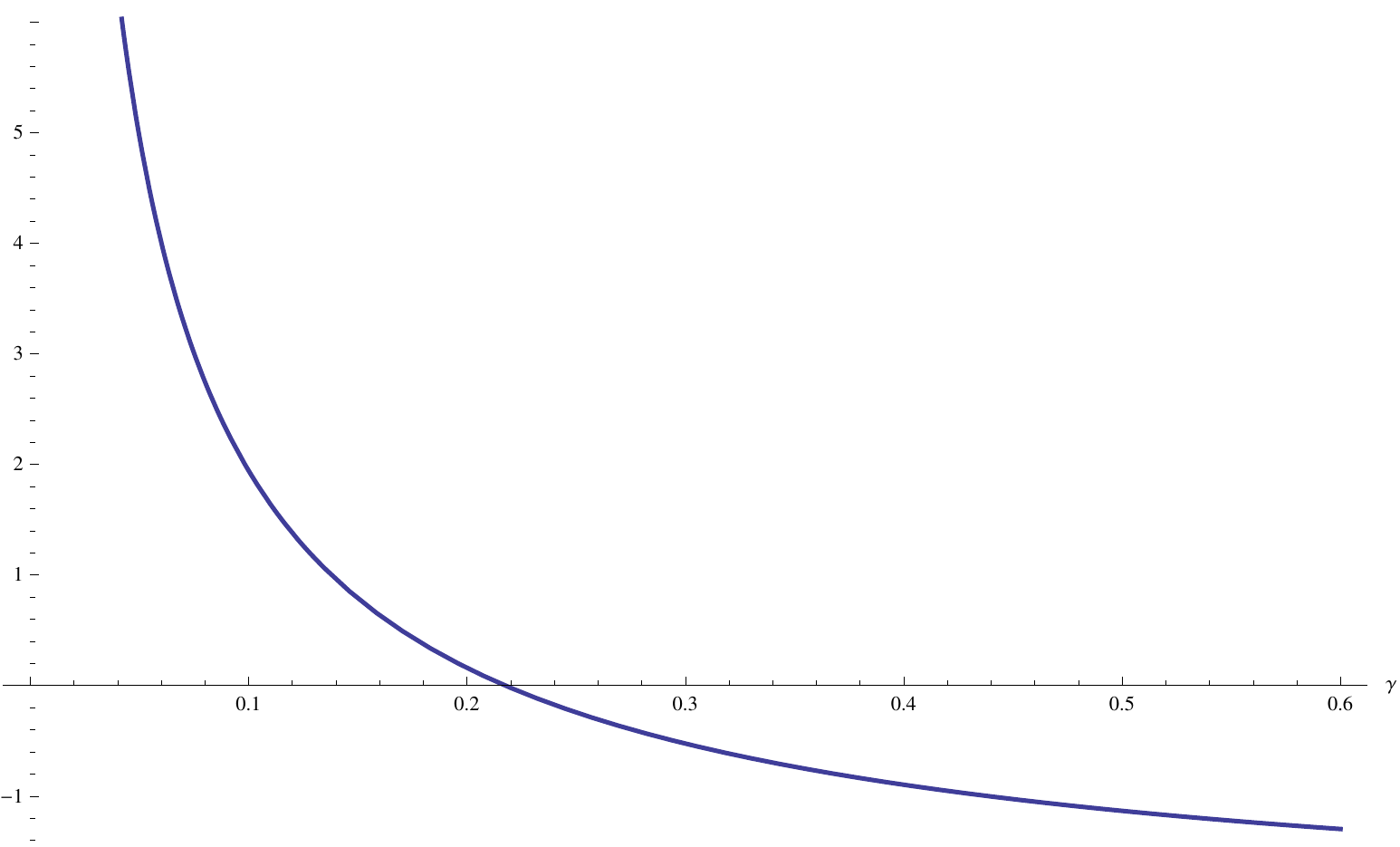}}}\hspace{6pt}
		\subfigure[Computation of $\gamma$ for $\mcl=2$ and $\overline{N}=2$]{
			\resizebox*{6.5cm}{!}{\includegraphics{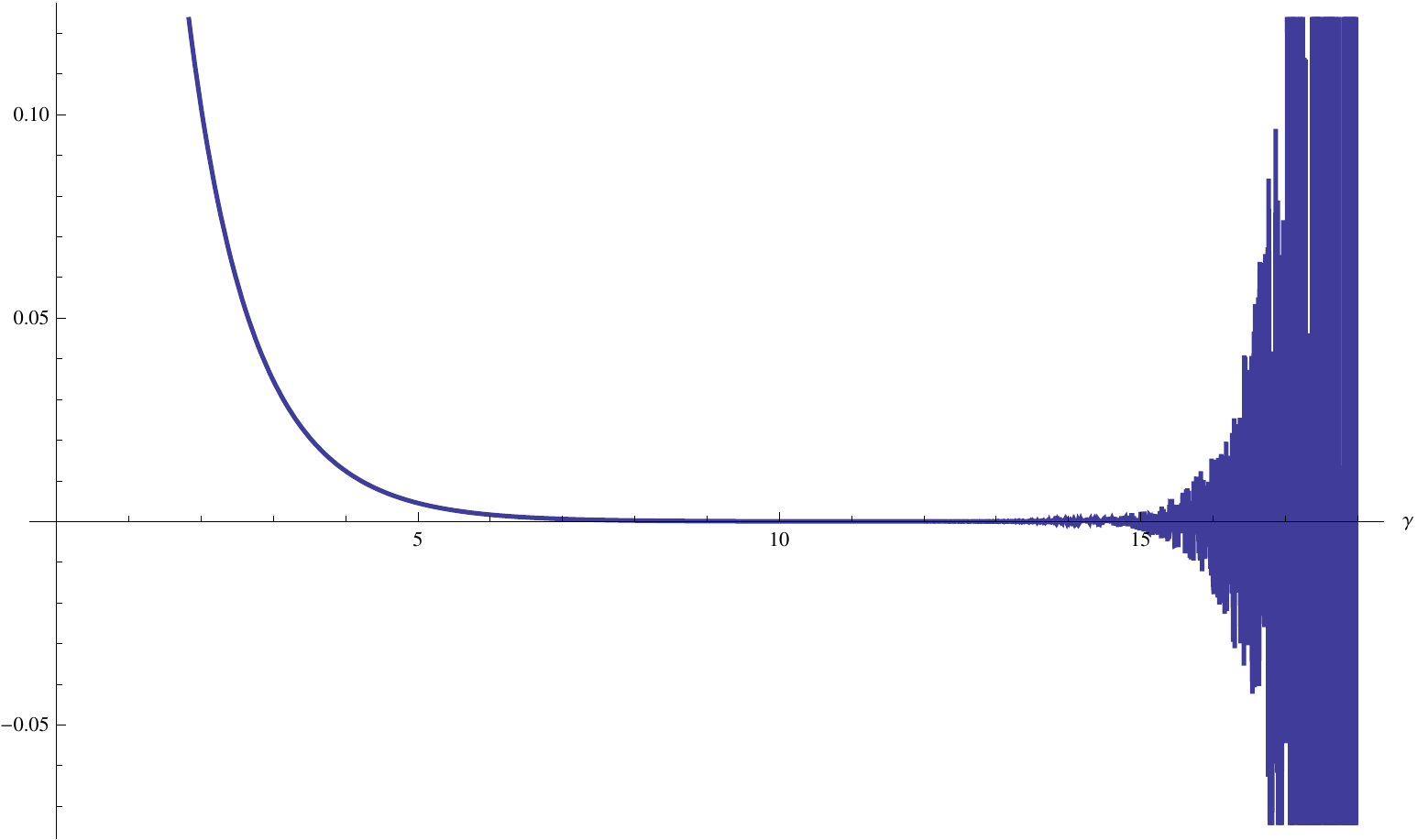}}}
		\caption{\label{fig:GammaNST2} Example curves for the root determination for the computation of gamma.}
		\label{fig:GammaNST}
	\end{center}
\end{figure}

The accuracy of this result depends on the fixed parameters $\mcl$ and $\overline{N}$. In the case $\mcl < \overline{N}$ the function has a unique root. If this condition is not fulfilled, the determination proves difficult. Since the model fulfills the condition $\mcl = \mincn \leq \overline{N}$, the only case left to consider is $\mcl = \overline{N}$. Then, the function approaches zero and diverges from some point onwards, as shown in Figure \ref{fig:GammaNST2}.

From equation \ref{sm:WN}, the ratio of the first two $W_N$ can be derived:
\begin{equation}\label{sm:case-only-one-cn}
	\frac{W_{\mcl}}{W_{\mcl+1}}=\frac{e^{-\gamma \mcl}}{\mcl}\frac{\mcl+1}{e^{-\gamma (\mcl+1)}} = \frac{\mcl+1}{\mcl}e^{\gamma}
\end{equation}

In the case $\mcl = \overline{N}$, this ratio (and therefore $\gamma$) would have to approach infinity. 
However, this case would also lead to only one non-zero class probability $\newP_1 = 1$, where an approximation is not meaningful anyways.

If it is still necessary to evaluate the formula for $\mcl = \overline{N}$, it appears to be a good solution to locate the root where there are still no oscillations. For instance, the exact root does not need to be determined and  there are no rigid requirements towards accuracy.
Thus, it is e.g. possible to consider only the first 8 decimal places to obtain a good root approximation.

In any case, the chosen algorithm needs to be very robust since we have a strongly non-linear function.

However, if $\gamma$ (or a similar quantity like $\exp(-\gamma)$) is used as an optimization parameter, it is usually not necessary to compute $\gamma$ from a given $\overline{N}$.

\subsection{Optimal $\mathbf{\mcl}$ for a given $\mathbf{\gamma}$.}\label{sec:opt_n0_for_given_gamma}
In Section \ref{sec:BerechnungPi} it was shown that, for a given $\gamma$, the unnormalized probabilities for all possible $\mcl$ are sums over the same sequence of values $\exp(-\gamma k)/k^{2}$, but with different starting indices.
This also means that they are all part of the same cumulative sum.
By computing this cumulative sum once for a given $\gamma$, it is possible to very quickly obtain the individual distributions $\mathbb{P}(\gamma,\mcl)$ for each $\mcl$. In this way, assuming a plausible upper limit $\mcl^{\text{max}}$, the optimal $\mcl$ can be obtained simply by evaluating every possible value.

Tests on real examples (Section \ref{sec:Beispiele}) showed that the function $Error(\mcl)$ for a given $\gamma$ can have one of two forms: Either the minimum is reached relatively quickly ($\mcl$<10) or $Error(\mcl)$ slowly descends towards a lower bound for large $\mcl$. To avoid the arbitrary result $\mcl=\mcl^{\text{max}}$ in the latter case, one can instead choose the optimal $\mcl$ by defining a \enquote{convergence} criterion:

\textbf{Definition: \textit{Optimal $\mathbf{\mcl}$ for a given $\mathbf{\gamma}$}.} For a given parameter $\gamma$, source distribution $\mathbb{D}$ and $Accuracy^{\mcl}>0$, the optimal $\mcl$ is defined as the largest $\mcl$ below the global minimum of the $Error$-function for which:  $|Error(\mcl-1)-Error(\mcl)|>Accuracy^{\mcl}$ or, if such an $\mcl$ does not exist, 1.

The value $Accuracy^{\mcl}$ should be chosen such that the altering of $\mcl$ has a negligible effect on the overall error. $\mcl^{\text{max}}$ should be set so that the convergence is always reached before $\mcl^{\text{max}}$.

For this work, we choose $Accuracy^{\mcl}=0.0001$, a value roughly 500 times smaller than the average error, and $\mcl^{\text{max}}=200$. However, this choice depends on the bounds for the parameter $\gamma$ - for values close to $0$ or $1$, more extreme settings might be necessary.


For large $\mcl$, small changes in $\mcl$ have a limited influence on $\mathbb{P}$. Therefore, these choices have little effect on the resulting approximation, and differences in $\mcl$ values should always first be investigated by considering the respective $Error(\mcl)$-graphs.


\subsection{Summary of the approximation algorithm}
\label{sec:ApproAlgorithmus}

To obtain a summarized overview of the approximation algorithm, the various steps and their connection will now be outlined (Algorithm~\ref{fig:Approximationsverfahren}).

\begin{algorithm}[!ht]
	\footnotesize
	\label{fig:Approximationsverfahren}
	\SetKwInOut{Input}{input}\SetKwInOut{Output}{output}
	\SetFuncSty{textbf}
	\SetKw{KwBy}{by}
	\let\oldnl\nl
	\newcommand{\nonl}{\renewcommand{\nl}{\let\nl\oldnl}}
	\SetKwFunction{myalg}{ApproximateDistribution}
	\DontPrintSemicolon
	\nonl\myalg{$\mathbb{D}$}\\
	\Input{An arbitrary distribution $\mathbb{D}$ with $m$ elements and $D_{n}$ denoting the value at position $n$ for $n=1,...,m$}
	\Output{An approximated probability distribution $\mathbb{P}$ from distribution $\mathbb{D}$ with $\newP_{n}$ denoting the value at position $n$ for $n=1,...,m$}
	\BlankLine
	\Begin{
		$sortDescending(\mathbb{D})$\;
		$normalize(\mathbb{D})$\;
		$Accuracy^{\mcl} \longleftarrow 0.0001$  \tabto{8cm} [Section \ref{sec:opt_n0_for_given_gamma}]\;
		$\mcl^{\text{max}} \longleftarrow 200$  \tabto{8cm} [Section \ref{sec:opt_n0_for_given_gamma}]\;
		$Error^{\text{min}} \longleftarrow \infty$\;
		$\mcl^{\text{optimal}} \longleftarrow -1$\;
		$z^{\text{optimal}} \longleftarrow -1$\;
		\For{$z \longleftarrow 0.0005$ \KwTo $0.9995$ \KwBy $0.0005$}{
			$Errors \longleftarrow [0,\dots,0]$\;
			\For{$\mcl \longleftarrow 1$ \KwTo $\mcl^{\text{max}}$ \KwBy $1$}{
				$\mathbb{P} \longleftarrow computeProbabilities(z,\mcl)$  \tabto{7cm} [Section \ref{sec:BerechnungPi}]\;
				$Errors[\mcl] \longleftarrow calculateError(\mathbb{D},\mathbb{P})$ \tabto{7cm} [Section \ref{sec:GütePi}]\;
			}
			$\mcl, Error \longleftarrow findOptimal\mcl(Errors)$  \tabto{7.5cm} [Section \ref{sec:opt_n0_for_given_gamma}]\;
			\If{$Error < Error^{\text{min}}$}{
				$Error^{\text{min}} \longleftarrow Error$\;
				$\mcl^{\text{optimal}} \longleftarrow \mcl$\;
				$z^{\text{optimal}} \longleftarrow z$\;
			}
		}
		\Return $computeProbabilities(x^{\text{optimal}},\mcl^{\text{optimal}})$
	}
	\caption{Approximate a distribution}
	\normalsize
\end{algorithm}
As the starting point for the approximation, an arbitrary frequency distribution of features is required. Since it deals with classification features, the order of the individual characteristics does not matter. However, the algorithm needs a distribution sorted by frequency in descending order to approximate properly.
Therefore, frequencies have to be preprocessed, i.e. at first they need to be sorted in descending order and then rearranged into a relative frequency distribution. The relative frequency of a feature is here calculated using equation \ref{sm:relHaeufigkeit}. Now, the aim is to approximate a distribution, which represents the frequency distribution that emerged during the preprocessing step as well as possible. This requires the determination of values for the parameters $\mcl$ and $\overline{N}$ such that the normalized error of the approximated distribution, with respect to the initial distribution, is minimal.

In practice, it is useful to optimize $\gamma$ and $\mcl$ and then calculate the optimal $\overline{N}$ from these values.
For the optimization itself, one option is to initialize the two parameters with random values and afterwards improve them step by step with a multidimensional optimization algorithm.


Since, for a certain $\gamma$, all (unnormalized) probabilities $\newP_n$  are part of the same partial sum, the transcendent function $\Phi$ only has to be evaluated once to be able to compute all $\newP_n$ for all possible $\mcl$.
This turns the optimization effectively into a one-parametric fit (as the second parameter $\mcl$ can always be chosen optimal, Section \ref{sec:opt_n0_for_given_gamma}).

For the optimization of $\gamma$, it proved useful to iterate over the $z := e^{-\gamma}$:
\begin{equation}
	z := e^{-\gamma} \in (0,1) \qquad \text{ and } \qquad \gamma = -\ln z
\end{equation}

Tests on real examples showed that the function $Error(z)$ still varies greatly and may contain many local minima. At the same time, it is continuous. 
\par
Because of this, instead of a conventional optimization algorithm, for the initial testing done in this work we instead compute the error for each of a given set of z-values and use the one with the lowest error (Algorithm \ref{fig:Approximationsverfahren})\footnote{A python script that implements this method is included in the ancillary files of this publication.}.
This has the benefit that the global minimum is always roughly found. The downside is that the accuracy of the parameter $\gamma$ is fairly low. For the approximations in this work, an accuracy of a few digits is sufficient. However, this problem could also be solved by running the same algorithm recursively in a small range around the previously found z until the desired accuracy is reached.




In applications where efficiency or high accuracy is important, it would be more practical to apply a specialized one-parametric minimization algorithm to find the optimal $\gamma$, instead of the simple graphical method chosen here.

Once the optimum is found, the resulting distribution represents the approximation result and has many practical applications. In Section \ref{sec:Beispiele}, some of these practical examples will be introduced.

\setlength{\extrarowheight}{1ex}
\renewcommand{\arraystretch}{0.6}
\newcommand{\chartheight}{7cm}


\section{Evaluation and testing}
\label{sec:Beispiele}
\subsection{Evaluation methods}
\label{sec:eval-method}

\subsubsection{Significance testing}
An approximation of real world examples will never be perfect. To distinguish whether this deviation is statistically significant (i.e. can not be explained by statistical fluctuations alone), it is necessary to compute the expected error distribution $p(E)$ for a given probability distribution $\mathbb{P}=\newP_1,\dots,\newP_m$.

$p(E)$ will depend on theoretical class frequencies $\newP_n$ as well as the number of total elements $|\mathbb{M}|$.

The class frequencies $\newP_n$ given in Section \ref{sec:erweiterungModell} represent the expected values that the relative frequencies defined in Section \ref{sec:grundlagenStatistik} approach when the size of the base set $|\mathbb{M}|$ goes to infinity. However, in cases with a small number of elements, the model might produce frequency distributions which violate the demanded descending order. In that case, the two respective classes would have to be switched. This effect introduces a bias for small numbers of elements.

One method for the estimation of $p(E)$ despite this bias is to randomly generate a large number of frequency distributions from $\mathbb{P}$ and then count the number of times the error was below the chosen value.

The probability that, if the model hypothesis is right, the error is at least as large as the observed error $Error(\mathbb{D},\mathbb{P})$ is given by the $p$-value:
\begin{equation}
p := p\left(E\geq Error(\mathbb{D},\mathbb{P})\right)
\end{equation}

If this value is very small, the $Error(\mathbb{D},\mathbb{P})$ cannot be explained by statistical fluctuations alone, and a more complex model is needed to truly explain the observed data.

This approach, however, is only possible if the number of elements $|\mathbb{M}|$ is known and the uncertainty of the source distribution is plausibly dominated by these statistical fluctuations and not, for example, measurement uncertainty.

\subsubsection{Comparison to alternative distributions}\label{sec:alt}
Another approach to evaluate the quality of the obtained approximation are comparisons with alternative distributions. Two of these will be discussed in the following section.

\paragraph{Zipf's law}
\ \par
In Section \ref{sec:intro-zipf}, Zipf's law (equation \ref{eq:simpleZipf}) was discussed as an alternative distribution that could be applied to classification features.

Interestingly, the frequencies $f_n$ in equation \ref{eq:simpleZipf} have the same form as $W_N$ for $\gamma=0$. However, in that case the infinite series over $W_N$ diverges and thus cannot be normalized.

The difference between Zipf's law for two neighboring classes in the case $s = 1$ is:
\begin{equation}
f_n-f_{n+1} \propto \frac{1}{n}-\frac{1}{n+1} = \frac{n+1-n}{n \cdot (n+1)} = \frac{1}{n \cdot (n+1)}
\end{equation}

This can be compared with equation \ref{eq:recursion-diff}:
\begin{equation}
\newP_n - \newP_{n+1} 
\propto \frac{e^{-\gamma (n+\mcl-1)}}{(n+\mcl-1)^2}
\end{equation}

The difference between these formulas disappears for large $n$ and $\gamma\to0$.

From these similarities in form, it is plausible that distributions described well by a Zipf-distribution can also be approximated well with the described model.

\paragraph{Exponential distribution}
\ \par

The other distribution chosen for comparison in this work is the exponential distribution:
\begin{equation}
f_n \propto e^{-a\cdot n} \qquad a>0
\end{equation}

It occurs when the probability of a certain quantity de- or increasing is directly proportional to the quantity itself (e.g. radioactive decay or bacterial growth).
Section \ref{sec:joint-ent-instead} in the appendix contains a proof that this distribution is equivalent to a maximized joint entropy of the model mentioned in Section \ref{sec:entropieVerteilung}.

\subsection{Fitting random data}
After normalizing and sorting the input frequency distribution, most sufficiently varied data will have roughly the same shape as the $\newP_{n}$ probabilities and thus can also be fit by the approximation algorithm to produce >>good-looking<< error values.

\begin{figure}[htbp!]
	\centering
		\includegraphics[width=\textwidth]{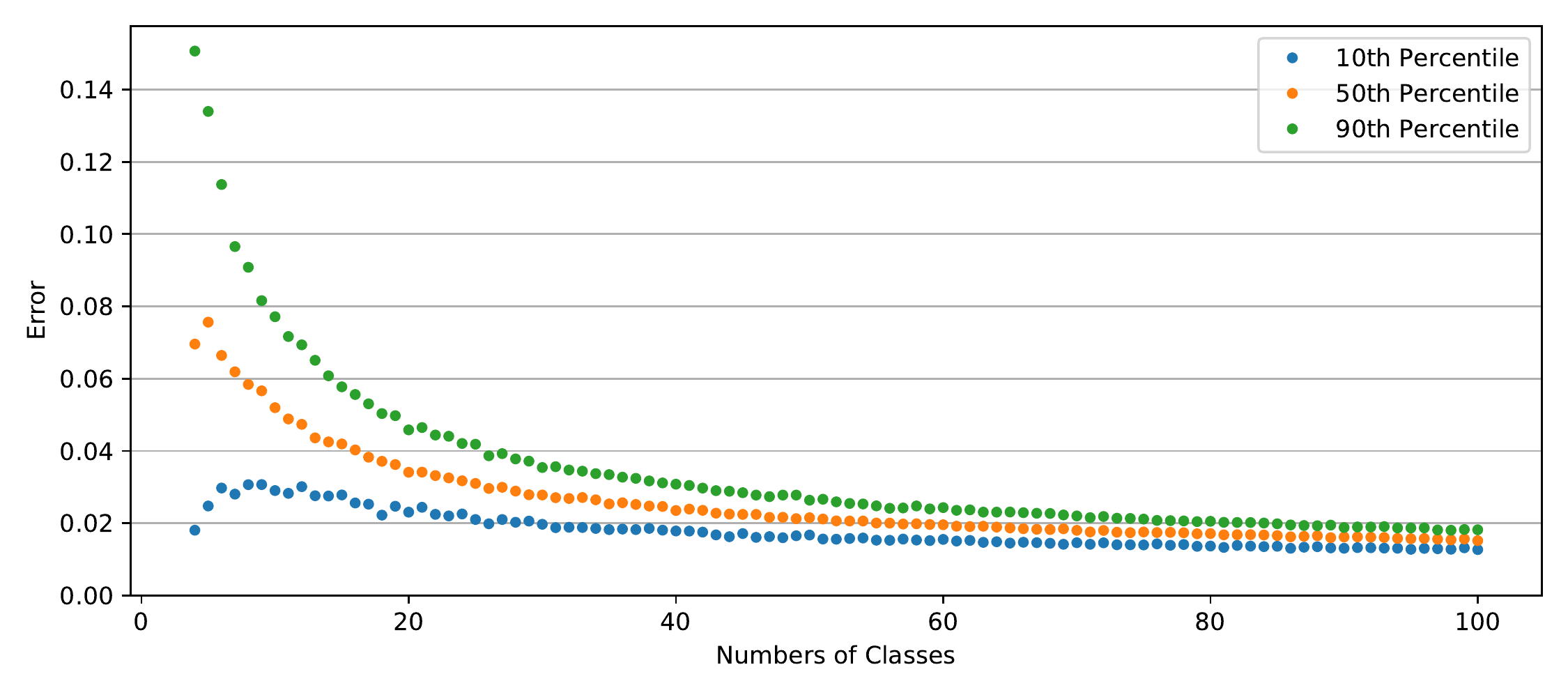}
		\caption{Results of approximating 500 ensembles of multiple-element samples drawn from a uniform distribution $[0,1]$ per class number.}
		\label{fig:rand-dat-perc}
\end{figure}

This effect can be investigated by applying the methods to randomly generated data. 
Figure \ref{fig:rand-dat-perc} shows the result of approximating ensembles of multiple-element samples drawn from a uniform distribution $[0,1]$.

The graph shows that the algorithm can approximate a wide variety of distributions and class numbers to a high degree (below $4\%$).
At the same time, the ability to fit random data with low error-values limits the amount of insight that can be drawn about a dataset from a close approximation alone.

\subsection{Real-world examples}

In the following section, the approximation algorithm described in the previous sections will be applied to real-world examples from various different domains. An overview over the chosen examples is provided in Table \ref{tab:examples}.

\begin{table}[!htbp]
	\centering
	\begin{minipage}{6in}
		\centering
\let\thempfootnote\thefootnote
\caption{Overview over the real-world examples}\label{tab:examples}
\vspace{3pt}
\setcounter{footnote}{1}
\footnotetext{For large numbers of elements, it is not possible to calculate meaningful p-values with the proposed method because other influences on the uncertainty outweigh the pure statistical fluctuations.}
\setcounter{footnote}{2}
\footnotetext{For distributions which have units associated with them, there is no clear number of elements.}
\setcounter{footnote}{3}
\footnotetext{A low p-value indicates that the deviation between model and data is statistically significant.}
\setcounter{footnote}{4}
\footnotetext{This row will be ignored during the calculations for Table \ref{tab:correl} and Table \ref{tab:avgerr}, since it is just a modification of the other internet host statistic.}
\setcounter{footnote}{5}
\footnotetext{Frequency distribution recorded with the system \textit{iSuite} described in chapter 8 of \cite{semantic-index-2019}.}

\footnotesize{
\begin{tabular}{lccccccc}
\toprule
Fig & Base Set $\mathbb{M}$ & $|\mathbb{M}|$ & Classes & $m$ & Error & p\footnotemark[3] & Ref \\\midrule

\midrule
Fig. \ref{fig:dual_chronic_pain_patients} & chronic pain cases &  & pain category & 24 & 0.039 &  & \footnotemark[5] \\
Fig. \ref{fig:dual_letter_frequency_en} & characters in english corpus & -\footnotemark[1] & english letters & 26 & 0.033 & - & \cite{LetterFreqENG} \\
Fig. \ref{fig:dual_countries_area_top_30} & land surface area & -\footnotemark[1] & countries & 30 & 0.048 & - & \cite{cia-worldfact-14} \\
Fig. \ref{fig:dual_gold_holdings_top_40_2015} & gold mass owned by countries & -\footnotemark[1] & countries & 40 & 0.046 & - & \cite{WBSP16} \\
Fig. \ref{fig:dual_countries_internet_hosts_top_40_2012} & internet hosts & many\footnotemark[2] & countries & 40 & 0.122 & - & \cite{cia-worldfact-14} \\
Fig. \ref{fig:dual_countries_internet_hosts_non_us_top_39_2012} & internet hosts outside the US & many\footnotemark[2] & countries & 39 & 0.047\footnotemark[4] & - & \cite{cia-worldfact-14} \\
Fig. \ref{fig:dual_counterfeit_euro_coins_2013} & counterfeit coins (\euro{}) & 52063 & coin values & 3 & 0.008 & 0.001 & \cite{WBSP11} \\
Fig. \ref{fig:dual_german_internet_connection_2014} & polled households & 33000 & connection types & 4 & 0.022 & 0 & \cite{WBSP7} \\
Fig. \ref{fig:dual_environmental_protection_employees_ger_2012} & employees in env. protection & many\footnotemark[2] & federal states & 16 & 0.029 & - & \cite{WBSP3} \\
Fig. \ref{fig:dual_open_job_positons_industry_branch_2014_q1} & open job positions in germany & many\footnotemark[2] & industry branches & 8 & 0.031 & - & \cite{WBSP26} \\
Fig. \ref{fig:dual_open_job_positons_occ_area_2014} & open job positions in Germany & many\footnotemark[2] & occup. areas & 19 & 0.034 & - & \cite{WBSP25} \\
Fig. \ref{fig:dual_bugs_linux_kernel_2014} & bugs in the linux kernel & 2952 & components & 20 & 0.037 & 0 & \cite{WBSP5} \\
Fig. \ref{fig:dual_countries_railway_top_40_2006_to_2012} & total length of railway tracks & -\footnotemark[1] & countries & 40 & 0.042 & - & \cite{cia-worldfact-14} \\
Fig. \ref{fig:dual_letter_frequency_de} & characters in german corpus & -\footnotemark[1] & german letters & 30 & 0.043 & - & \cite{BSPBuchstabenDS2005} \\
Fig. \ref{fig:dual_gdp_oecd_members_2012} & total OECD GDP & -\footnotemark[1] & countries & 34 & 0.047 & - & \cite{WBSP15} \\
Fig. \ref{fig:dual_countries_internet_users_top40_2009} & internet users & many\footnotemark[2] & countries & 40 & 0.049 & - & \cite{cia-worldfact-14} \\
Fig. \ref{fig:dual_bugs_glibc_2014} & bugs in glibc & 781 & components & 18 & 0.059 & 0.003 & \cite{WBSP6} \\
Fig. \ref{fig:dual_countries_population_top_40_july_2014} & human population & many\footnotemark[2] & countries & 40 & 0.070 & - & \cite{cia-worldfact-14} \\
Fig. \ref{fig:dual_german_motor_vehicles_production_country_2014} & registered cars in Germany & many\footnotemark[2] & prod. countries & 11 & 0.071 & - & \cite{WBSP18} \\
Fig. \ref{fig:dual_euro_cash_2014} & cash in circulation (\euro{}) & -\footnotemark[1] & coin/note values & 8 & 0.093 & - & \cite{WBSP19} \\
Fig. \ref{fig:dual_counterfeit_euro_notes_2013} & counterfeit bank notes (\euro{}) & 38811 & note values & 7 & 0.148 & 0 & \cite{WBSP11} \\
Fig. \ref{fig:dual_wikipedia_popular_pages_top_40_jan_2015} & website requests to Wikipedia & many\footnotemark[2] & indiv. pages & 40 & 0.190 & - & \cite{WBSP29} \\

\bottomrule
\end{tabular}

}
	\end{minipage}
\end{table}

The computed p-values indicate that all errors are statistically significant, i.e. they cannot be explained purely by statistical fluctuations.
This is mostly expected, since such a generic approach cannot account for  all complex processes that produce the individual distributions.

\begin{figure}[!htbp]\flushright\csvtofigrot{newcharts/chronic_pain_patients.csv}{Classes}{2}{\tiny}\includegraphics[width=0.93\linewidth]{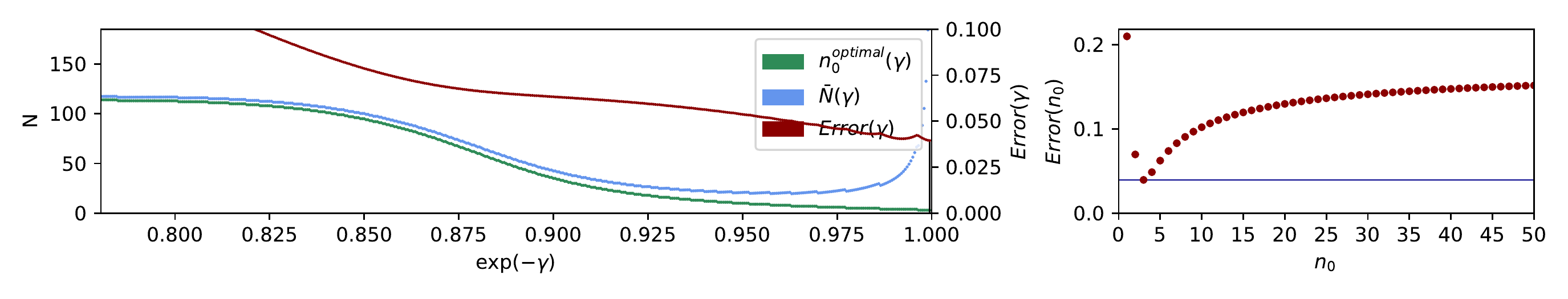}\caption{Symptoms of patients with chronic pain over a period of 7 days. Recorded with the system system \textit{iSuite} described in chapter 8 of \cite{semantic-index-2019}. Bottom: error behavior.}\label{fig:dual_chronic_pain_patients}\label{fig:indiv_a}\end{figure}

Figures \ref{fig:indiv_a} - \ref{fig:indiv_b} visualize some of the approximations\footnote{Additional examples are included in the appendix, Section \ref{sec:appendix-other-examples}.} and show the shape of $Error(\mathbb{D},\mathbb{P})$ as a function of $\exp(-\gamma)$ (left) as well as $\mcl$ at the optimal $\gamma$-value (right).

Figures \ref{fig:dual_letter_frequency_en}, \ref{fig:dual_countries_area_top_30} and \ref{fig:dual_gold_holdings_top_40_2015} illustrate the approximative power of the presented algorithm on completely different domains. The letter frequency of the English language clearly bears no relation to the world's countries with the largest area or most gold reserves, but the relation between the distributions is evident. In most cases, the approximated values approach the quality of the initial distribution with an error below $5\%$. 

\begin{figure}[!htbp]\flushright\csvtofig{newcharts/letter_frequency_en.csv}{Classes}{2}{\footnotesize}\includegraphics[width=0.93\linewidth]{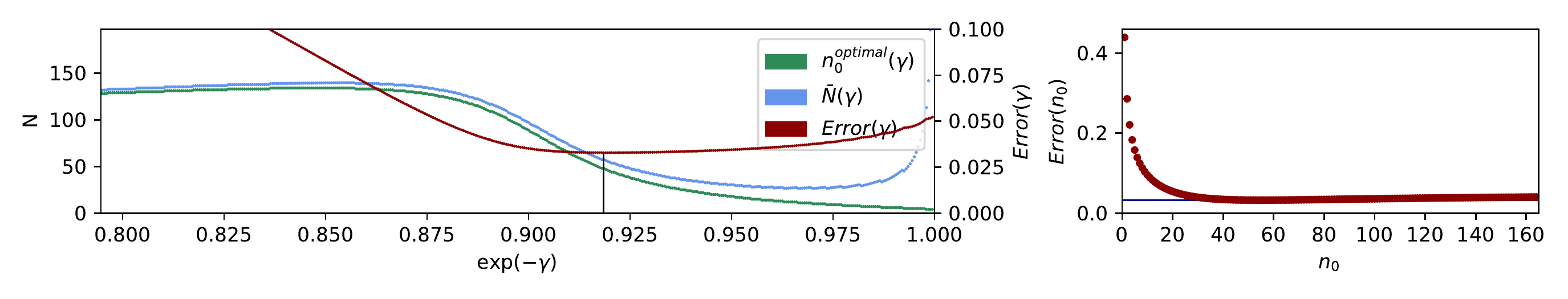}\caption{Frequency distribution of english letters \cite{LetterFreqENG} and error behavior.}\label{fig:dual_letter_frequency_en}\end{figure}
\begin{figure}[!htbp]\flushright\csvtofigrot{newcharts/countries_area_top_30.csv}{Classes}{1.5}{\tiny}\includegraphics[width=0.93\linewidth]{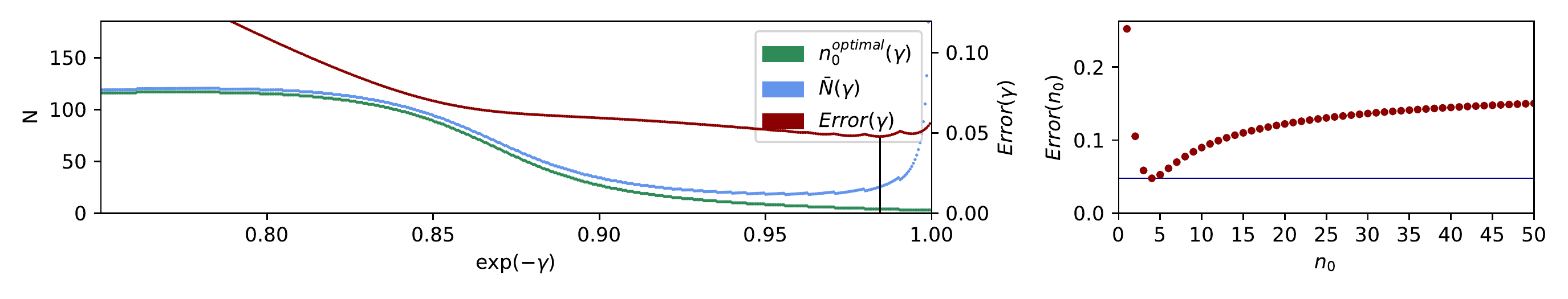}\caption{30 largest countries (2014) \cite{cia-worldfact-14} and error behavior.}\label{fig:dual_countries_area_top_30}\end{figure}
\begin{figure}[!htbp]\flushright\csvtofigrot{newcharts/gold_holdings_top_40_2015.csv}{Classes}{1.2}{\tiny}\includegraphics[width=0.93\linewidth]{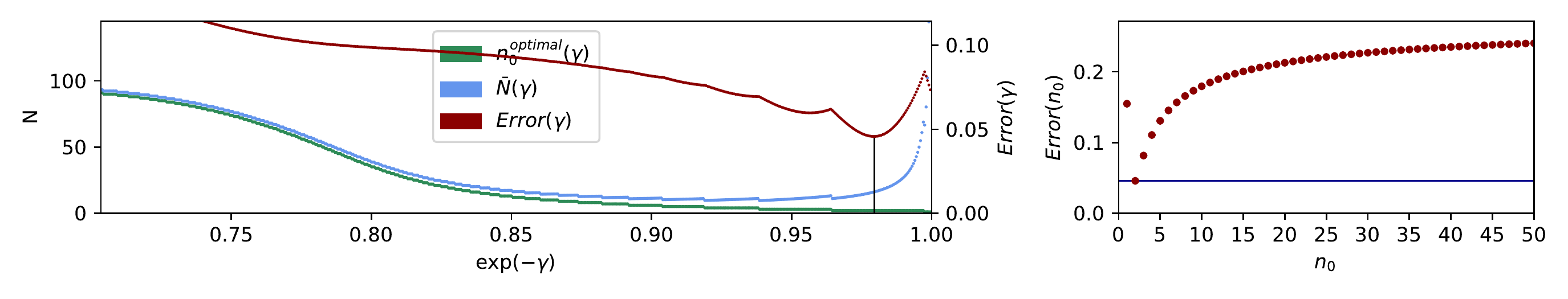}\caption{40 countries with the largest gold holdings in February 2015 \cite{WBSP16} and error behavior.}\label{fig:dual_gold_holdings_top_40_2015}\end{figure}

The correlation between different methods of analysis is given in Table \ref{tab:correl}.
\begin{table}[!htbp]
	\centering
	\caption{Pearson correlation coefficients}\label{tab:correl}
	\vspace{3pt}
	\begin{tabular}{lcccccc}
		\toprule
 & $\mcl$ & $Error_\text{Approx}$ & $Error_\text{Zipf}$ & $Error_\text{Exp}$ & $Error_{\text{Zipf}, s=1}$ & $Error_{\mcl=1,\mincn=2}$\\ \midrule
$m$ & -.33 & .06 & -.19 & .38 & -.39 & -.22\\
$\mcl$ &  & .21 & .56 & -.27 & .06 & -.37\\
$Error_\text{Approx}$ &  &  & .86 & .69 & .37 & -.08\\
$Error_\text{Zipf}$ &  &  &  & .36 & .21 & -.37\\
$Error_\text{Exp}$ &  &  &  &  & .26 & .20\\
$Error_{\text{Zipf}, s=1}$ &  &  &  &  &  & .77\\
		\bottomrule
	\end{tabular}
\end{table}

$Error_{\mcl=1,\mincn=2}$ is the approximation quality of the method presented in \cite{journals/eik/Voss69a}, where $\mincn$ was set to $2$ but no classes were excluded (cf. Section \ref{sec:min_nr_of_classes}).
The highest correlation appears between Zipf's law and the presented approximation algorithm. 
This supports the conclusion that distributions which suffice Zipf's law can be approximated and described especially well by the presented method. All methods except for the  $Error_{\mcl=1,\mincn=2}$ approximation are positively correlated.

Table \ref{tab:avgerr} shows the average and median error of the different methods across the tested examples. Algorithm \ref{fig:Approximationsverfahren} and the one-parametric Zipf distribution are comparable in quality and clearly outperform the other methods.
The introduction of a free parameter $\mcl=\mincn$ represents a clear improvement over the original case $\mincn=2, \mcl=1$ that was considered in \cite{journals/eik/Voss69a}.
\begin{table}[!htbp]
	
	\centering
	\caption{Approximation quality of the different methods}\label{tab:avgerr}
	\vspace{3pt}
	\begin{tabular}{lccccc}
		\toprule
Method & \textbf{Algorithm \ref{fig:Approximationsverfahren}} & Zipf & Exponential & Zipf, $s=1$ & $\mcl=1,\mincn=2$\\ \midrule
Average $Error$ & $\mathbf{0.053}$ & 0.064 & 0.090 & 0.140 & 0.148\\
Median $Error$ & $\mathbf{0.046}$ & 0.057 & 0.092 & 0.116 & 0.098\\
continuous parameters & $\mathbf{1}$ & 1 & 1 & 0 & 1\\
discrete parameters & $\mathbf{1}$ & 0 & 0 & 0 & 0\\

		\bottomrule
	\end{tabular}
\end{table}




\begin{figure}[!htbp]\flushright\csvtofigrot{newcharts/countries_internet_hosts_top_40_2012.csv}{Classes}{1.2}{\tiny}\includegraphics[width=0.93\linewidth]{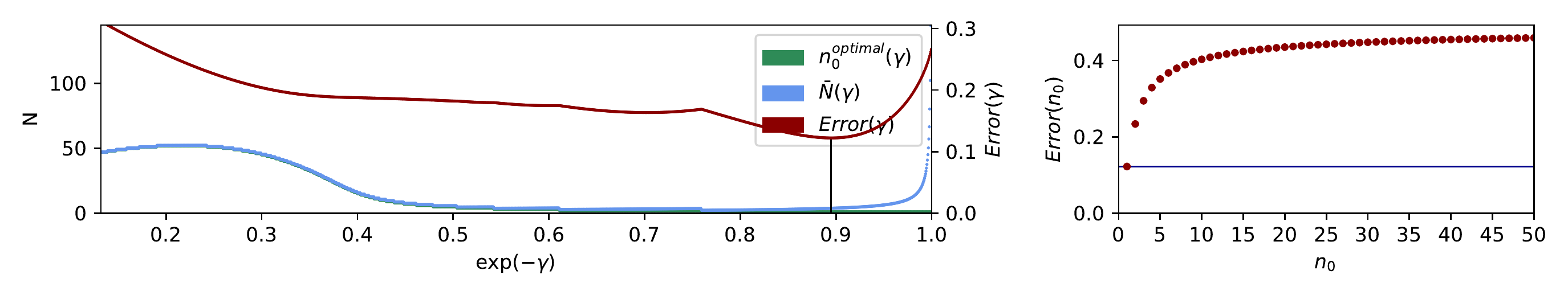}\caption{Number of internet hosts by country (2012, top 40) \cite{cia-worldfact-14} and error behavior.}\label{fig:dual_countries_internet_hosts_top_40_2012}\end{figure}
\begin{figure}[!htbp]\flushright\csvtofigrot{newcharts/countries_internet_hosts_non_us_top_39_2012.csv}{Classes}{1.2}{\tiny}\includegraphics[width=0.93\linewidth]{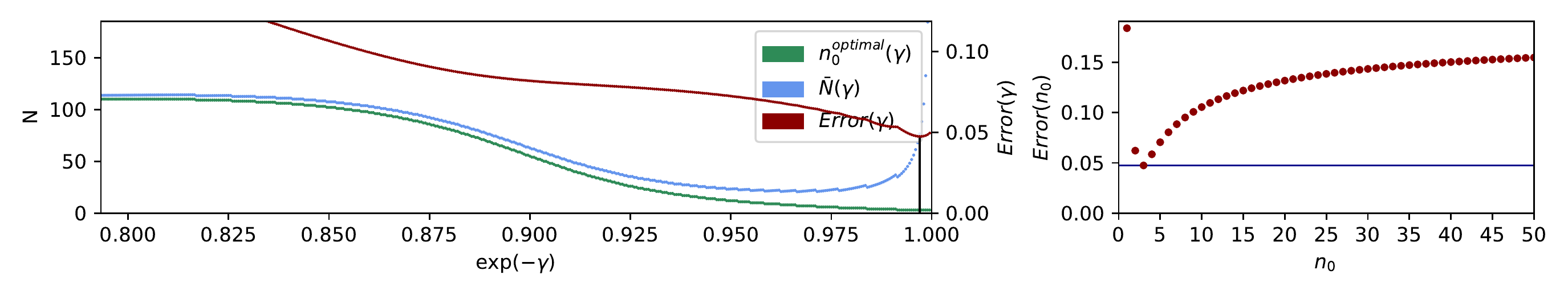}\caption{Number of internet hosts by country, excluding USA (2012, top 39) \cite{cia-worldfact-14} and error behavior.}\label{fig:dual_countries_internet_hosts_non_us_top_39_2012}\label{fig:indiv_b}\end{figure}

Figure \ref{fig:dual_countries_internet_hosts_top_40_2012} shows the number of internet hosts by country in 2012 \cite{cia-worldfact-14}. The large error value of over $12\%$ can be at least partially explained by the so-called King effect - the first class significantly outranks the rest of the distribution. If this first class is excluded the approximation error goes down to $4.7\%$ (Figure \ref{fig:dual_countries_internet_hosts_non_us_top_39_2012}).

\setlength{\extrarowheight}{0ex}
\renewcommand{\arraystretch}{1.0}

\section{Summary and outlook}
\label{sec:zusammenfassungUndAusblick}

The approximation algorithm for frequency distribution formulated in Section \ref{sec:ApproWKVerteilungen} has many applications and is able to describe and explain the attributes and behavior of complex systems. 
\par
The results of the numerous examples show an improvement and possible generalization compared to, e.g., Zipf's distribution. Through the addition of the new parameter $\mcl$, the principle of maximum entropy could be transferred to a statistical model of classification features that is comparable in approximation power to similar statistical laws. With the obtained result we developed not only a general and computable model but also an algorithm for the approximation of probability distributions of classification features. 
\par
If the approximations are examined without the background of the derived model, one could get the impression that it would probably be possible to provide a much simpler curve fit. However, the relations derived in this work are not a simple approximation, but a causal statistical modeling of relationships.
The presented examples show that the agreement between model and practice is guaranteed up to deviations of about 5\%.  
The approximations of many other real-world examples indicated a similar quality.
\par
The range of possible applications is considerable and warrants further research. In addition to simple approximation and smoothing scenarios the model could also, with some refinements, be applied in diverse regression analyses and forecasts, which were until now difficult to compute with other statistical models.
Another advantage is the comparability of approximations. Using the parameters $\mcl$ and $\overline{N}$, any number of values of a distribution can be calculated. This makes it possible to compare distributions with different class numbers by calculating the missing values of the distribution with the lower class number up to the distribution with the higher class number.
\par
We believe that this approach has potential and are looking forward to see what relevance the model will have in practice.

\appendix
\newpage
\section{Appendix}

\subsection{Computation of parameters $Z$ and $\alpha$ in $\newP_{n}$}
\label{sec:BerechnungZ}

It may be necessary to compute the partition function $Z$ and the normalization factor $1/\alpha$ to calculate the absolute probabilities of particular classes. With the knowledge gained in Section \ref{sec:BerechnungGamma}, the computation of $Z$ can also be simplified. By a rearrangement of the transcendental Lerch function analogous to equation \ref{appro:gammaherleitung2}, the infinite sum can be eliminated here as well:

\begin{equation}\label{appro:zherleitung}
\begin{aligned}
Z &\stackrel{\ref{sm:Z}}{=} \sum\limits_{N=\mcl}^{\infty} \frac{e^{-\gamma N}}{N} = \frac{e^{\gamma \mcl}}{e^{\gamma \mcl}} \cdot \sum\limits_{N=\mcl}^{\infty} \frac{e^{-\gamma N}}{N} \\
&= \frac{1}{e^{\gamma \mcl}} \cdot \sum\limits_{N=0}^{\infty} \frac{e^{-\gamma N}}{\mcl+N} \stackrel{\ref{appro:LerchPhi}}{=} e^{-\gamma \mcl} \cdot \Phi(e^{-\gamma}, 1, \mcl) \\
&\stackrel{\ref{appro:gammaherleitung2}}{=} e^{-\gamma \mcl} \cdot \tau(\mcl, \gamma) = e^{-\gamma \mcl} (-e^{\gamma(\mcl-1)}) \left( e^{\gamma} \cdot \ln(1-e^{-\gamma}) + \sum\limits_{k=0}^{\mcl-2} \frac{e^{-\gamma k}}{k+1} \right) \\
&= -e^{-\gamma} \left( e^{\gamma} \cdot \ln(1-e^{-\gamma}) + \sum\limits_{k=0}^{\mcl-2} \frac{e^{-\gamma k}}{k+1} \right) \\
&= -\ln(1-e^{-\gamma}) -e^{-\gamma} \sum\limits_{k=0}^{\mcl-2} \frac{e^{-\gamma k}}{k+1} = -\ln(1-e^{-\gamma}) -\sum\limits_{k=0}^{\mcl-2} \frac{e^{-\gamma(k+1)}}{k+1}\\
&= -\ln(1-e^{-\gamma}) -\sum\limits_{k=1}^{\mcl-1} \frac{e^{-\gamma k}}{k}
\end{aligned}
\end{equation}

Taking $\gamma$ into account, $Z$ can now also be computed directly and no evaluation of infinite series is required.

\begin{equation}\label{eq:calczfinal}
\boxed	{
	Z(\mcl, \gamma) = e^{-\gamma \mcl}\cdot\tau(\mcl, \gamma) =  -\ln(1-e^{-\gamma}) -\sum\limits_{k=1}^{\mcl-1} \frac{e^{-\gamma k}}{k}
}
\end{equation}

The computation of $\alpha$ can also be reformulated to make use of a previously implemented Lerch-transcendent:

\begin{equation}\label{appro:alpha}
\begin{aligned}
\alpha &\stackrel{\ref{sm:PnNormierung}}{=}
1-(\mcl-1)\sum\limits_{k=\mcl}^{\infty}W_{k} 
= 1- \frac{1}{Z} (\mcl-1) \sum\limits_{k=\mcl}^{\infty} \frac{e^{-\gamma k}}{k^{2}} \\
&= 1- \frac{1}{Z} (\mcl-1) \sum\limits_{k=0}^{\infty} \frac{e^{-\gamma k} \cdot e^{-\gamma \mcl}}{(k+\mcl)^{2}} \\
&= 1- \frac{1}{Z} (\mcl-1) \cdot e^{-\gamma \mcl} \cdot \Phi(e^{-\gamma}, 2, \mcl)
\end{aligned}
\end{equation}

The corresponding formula for the calculation of a concrete $\newP_{n}$ is: 

\begin{equation}\label{eq:calcpfinal}
	\newP_{n}(\mcl, Z, \gamma) = \frac {1} {Z} \cdot \frac {1} {\alpha} \cdot e^{-\gamma (n+\mcl-1)} \cdot \Phi(e^{-\gamma}, 2, \mcl + n - 1)
\end{equation}

\subsection{Short investigation of the results of a joint entropy maximation - exponential decline}\label{sec:joint-ent-instead}
In Section \ref{sec:entropieVerteilung}, the entropy measure $D=H(\distrN) - H(\distrn|\distrN)$ was chosen instead of the joint entropy $H = H(\distrN) + H(\distrn|\distrN)$. In the following section, we will examine the changes if the joint entropy had been maximized instead.

From equations \ref{sm:verbundentropie}  and \ref{sm:verbundentropiedifferenz}, the transformation $D \rightarrow H$ has the form:
\begin{equation}
\sum\limits_{N=\mcl}^{\infty} \sum\limits_{n=1}^{N} W(N,n) \left(\ln W(N,n) - 2 \cdot \ln W_N\right) \longrightarrow \sum\limits_{N=\mcl}^{\infty} \sum\limits_{n=1}^{N} W(N,n) \left(-\ln W(N,n)\right)
\end{equation}
Tracking this change in coefficients through the Lagrange differential (equation \ref{sm:lagrangedifferential1}) leads to:
\begin{equation}
0 \stackrel{}{=} 
-\ln W(N,n)-1
- \beta - \gamma N \stackrel{\ref{sm:WNn}}= - \ln W_N -\ln p_n^{(N)} -1 -\beta-\gamma N
\end{equation}
This again leads to the $n$-independent $p_n^{(N)}=\frac{1}{N}$ described by equation \ref{sm:GleichverteilungpnN}.

Derivation \ref{sm:HerleitungWN} transforms to:
\begin{equation}
\begin{aligned}
0 &= -\ln W_{N} - \ln{\frac{1}{N}} - \beta - \gamma N - 1\\
\ln W_{N} &= \ln N - \beta - \gamma N - 1\\
W_{N} &= e^{\ln N - \beta - \gamma N - 1} = N \cdot e^{-\beta-1} \cdot e^{-\gamma N} = \frac{1}{e^{\beta+1}} \cdot N \cdot e^{-\gamma N} = \frac{N \cdot e^{-\gamma N}}{Z}
\end{aligned}
\end{equation}

The derivations and changes in Sections \ref{sec:excludeclasses} and \ref{sec:erweiterungModell} are independent from the concrete form of $W_{N}$, so the formula \ref{sm:PnNormierung} still holds:

\begin{equation}
\newP_n = \frac{1}{\alpha} \sum\limits_{k=n+\mcl-1}^{\infty} \frac{1}{k} W_{k} \qquad \text{with} \quad \alpha = 1-(\mcl-1)\sum\limits_{k=\mcl}^{\infty}W_{k}
\end{equation}

It follows:
\begin{equation}
\newP_n = \frac{1}{\alpha \cdot Z} \sum\limits_{k=n+\mcl-1}^{\infty} e^{-\gamma k} \stackrel{\text{ Geometric series }}{=}\frac{e^{-\gamma(n+\mcl-2)}}{\alpha \cdot Z \cdot \left(e^{\gamma}-1\right)} \propto e^{-\gamma \cdot n}
\end{equation}
The parameter $\mcl$ is eliminated by the normalization.

Therefore, maximizing the joint entropy instead of the chosen difference $D$ would have lead to an exponential distribution $\newP_n \propto \exp(-\gamma n)$ with only 1 parameter $\gamma$.

This alternative model could be further investigated in connection with research areas where exponential distributions are commonly used.
\newpage
\subsection{Other examples}\label{sec:appendix-other-examples}

\begin{figure}[!htbp]\flushright\csvtofig{newcharts/counterfeit_euro_coins_2013.csv}{Classes}{9}{\footnotesize}\includegraphics[width=0.93\linewidth]{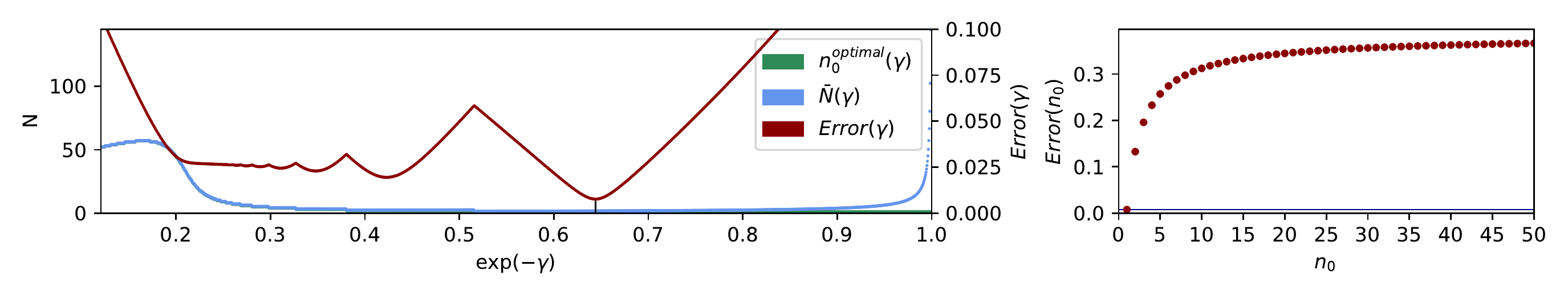}\caption{Counterfeit \euro{}-coins by coin value (2013) \cite{WBSP11} and error behavior.}\label{fig:dual_counterfeit_euro_coins_2013}\end{figure}
\begin{figure}[!htbp]\flushright\csvtofig{newcharts/german_internet_connection_2014.csv}{Classes}{7}{\footnotesize}\includegraphics[width=0.93\linewidth]{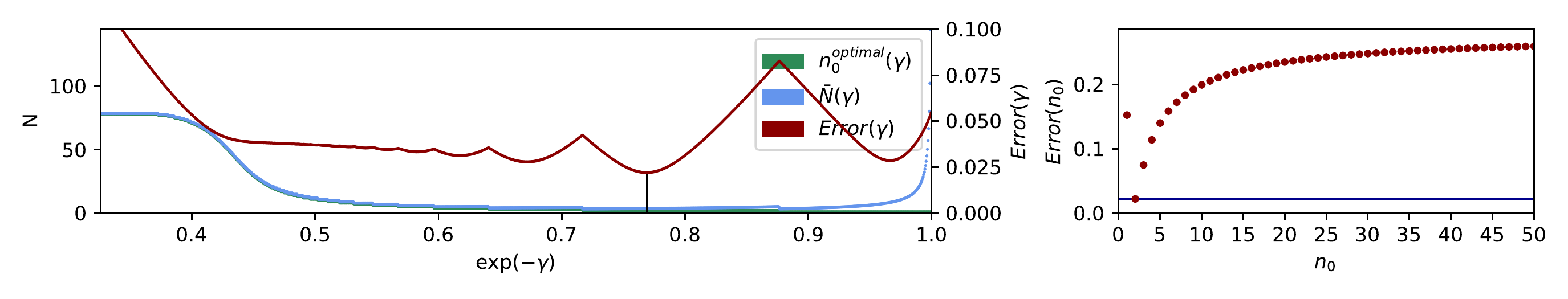}\caption{Type of internet connection of 33000 german households (2014) \cite{WBSP7} and error behavior.}\label{fig:dual_german_internet_connection_2014}\end{figure}
\begin{figure}[!htbp]\flushright\csvtofigrot{newcharts/environmental_protection_employees_ger_2012.csv}{Classes}{3}{\tiny}\includegraphics[width=0.93\linewidth]{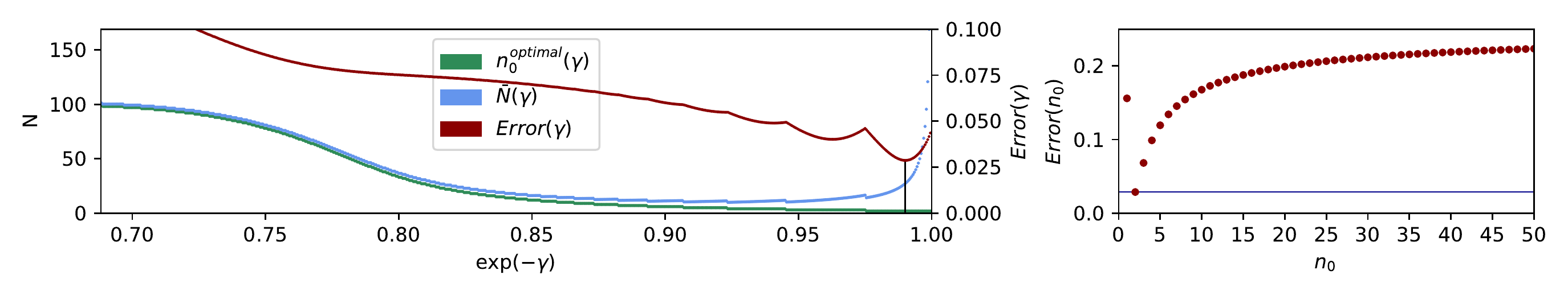}\caption{Number of employees in environmental protection, by federal state (Germany, 2012) \cite{WBSP3} and error behavior.}\label{fig:dual_environmental_protection_employees_ger_2012}\end{figure}
\begin{figure}[!htbp]\flushright\csvtofigrot{newcharts/open_job_positons_industry_branch_2014_q1.csv}{Classes}{4}{\tiny}\includegraphics[width=0.93\linewidth]{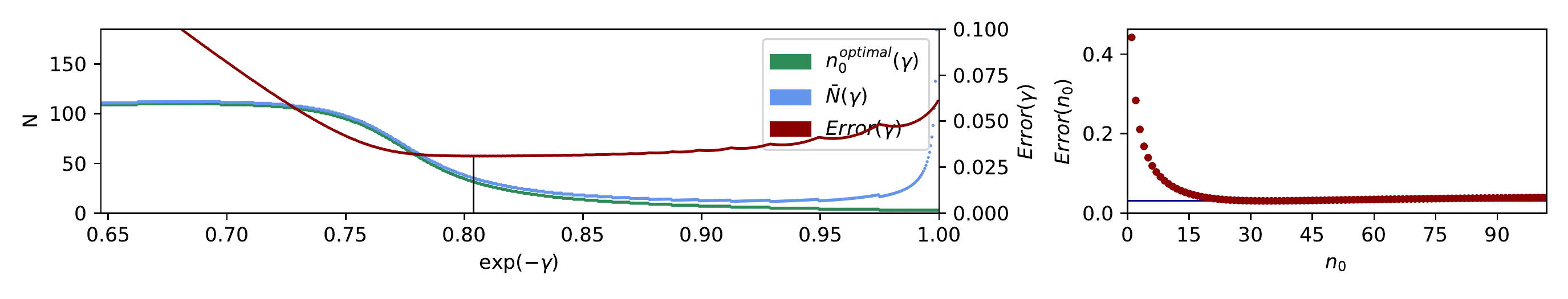}\caption{Industry branches with the most open job positions in Germany (2014, 1st quarter) \cite{WBSP26} and error behavior.}\label{fig:dual_open_job_positons_industry_branch_2014_q1}\end{figure}
\begin{figure}[!htbp]\flushright\csvtofigrot{newcharts/open_job_positons_occ_area_2014.csv}{Classes}{2.2}{\tiny}\includegraphics[width=0.93\linewidth]{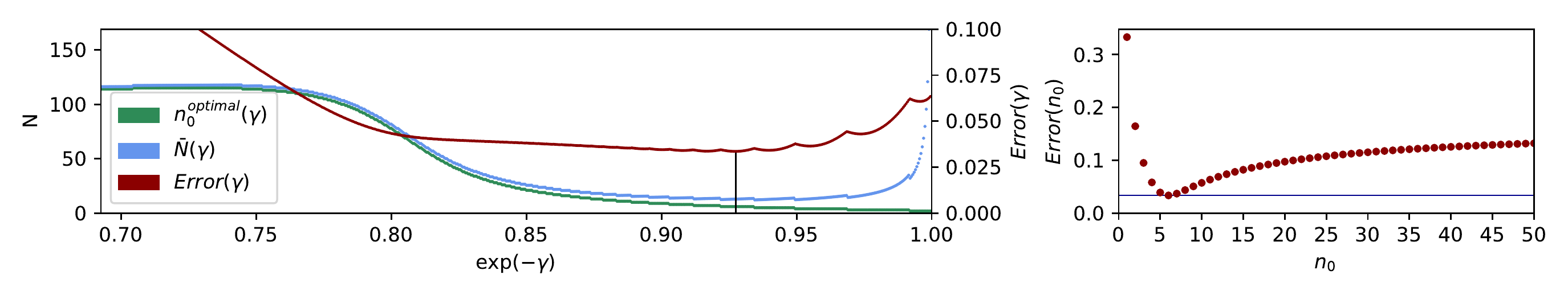}\caption{Open job positions in Germany in 2014, by occupational area \cite{WBSP25} and error behavior.}\label{fig:dual_open_job_positons_occ_area_2014}\end{figure}
\begin{figure}[!htbp]\flushright\csvtofigrot{newcharts/bugs_linux_kernel_2014.csv}{Classes}{2}{\tiny}\includegraphics[width=0.93\linewidth]{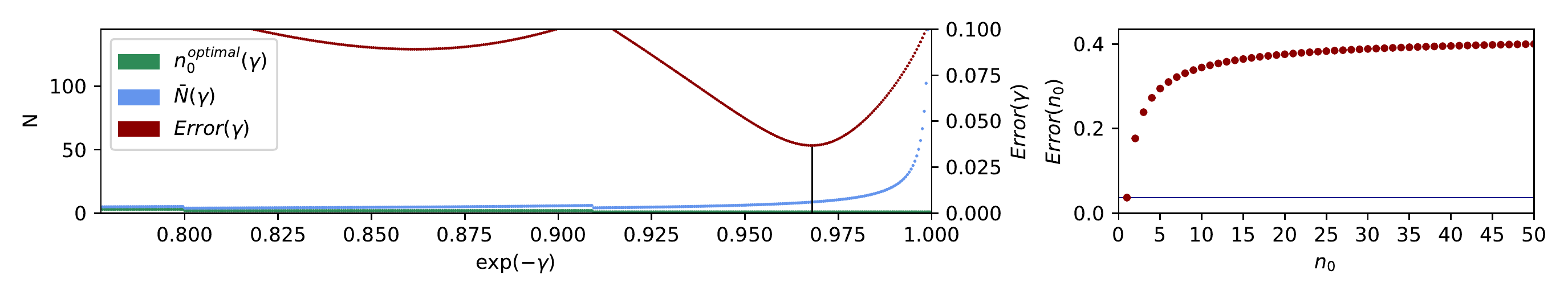}\caption{Number of bugs in the Linux kernel, by component (2014) \cite{WBSP5} and error behavior.}\label{fig:dual_bugs_linux_kernel_2014}\end{figure}
\begin{figure}[!htbp]\flushright\csvtofigrot{newcharts/countries_railway_top_40_2006_to_2012.csv}{Classes}{1.5}{\tiny}\includegraphics[width=0.93\linewidth]{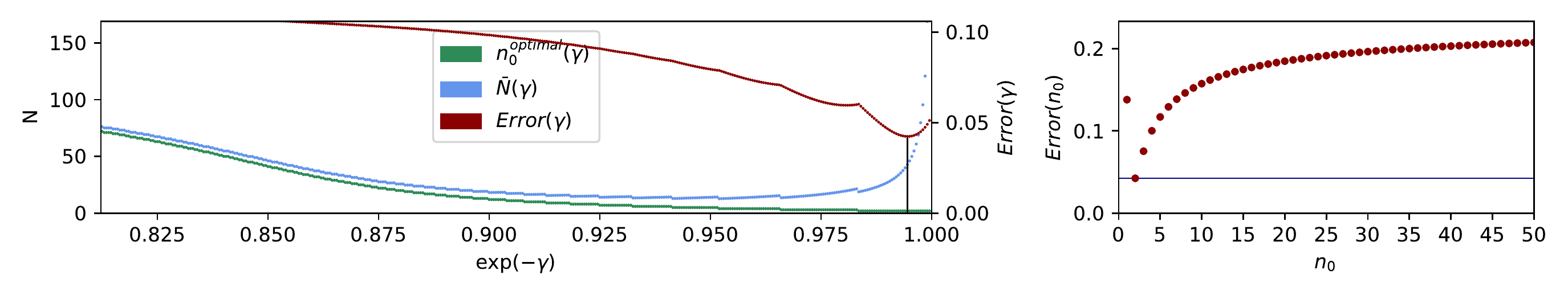}\caption{Total route length of the railway network by country (top 40, 2006 - 2012) \cite{cia-worldfact-14} and error behavior.}\label{fig:dual_countries_railway_top_40_2006_to_2012}\end{figure}
\begin{figure}[!htbp]\flushright\csvtofig{newcharts/letter_frequency_de.csv}{Classes}{1.5}{\footnotesize}\includegraphics[width=0.93\linewidth]{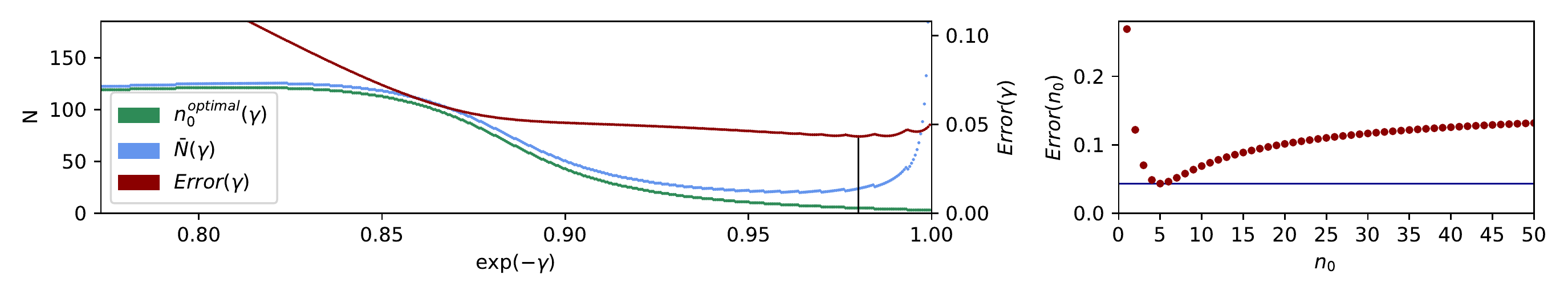}\caption{Frequency distribution of german letters \cite{BSPBuchstabenDS2005} and error behavior.}\label{fig:dual_letter_frequency_de}\end{figure}
\begin{figure}[!htbp]\flushright\csvtofigrot{newcharts/gdp_oecd_members_2012.csv}{Classes}{1.5}{\tiny}\includegraphics[width=0.93\linewidth]{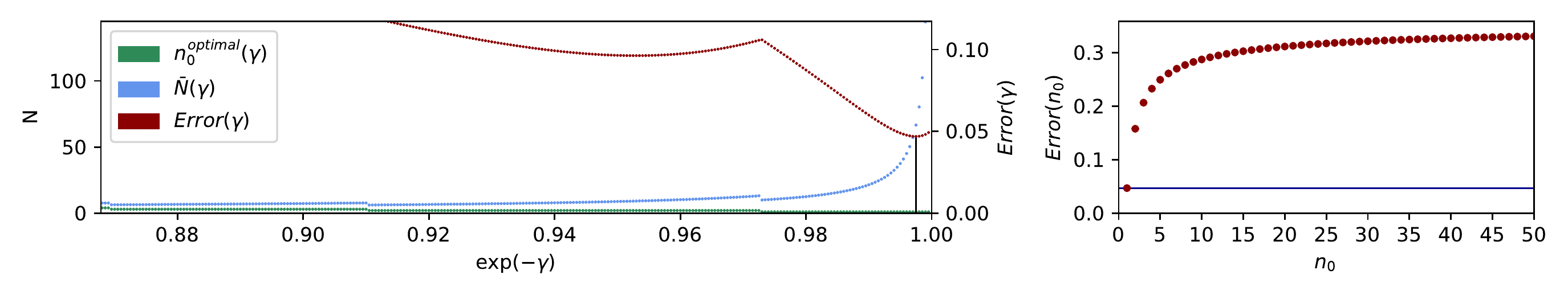}\caption{GDP of the OECD member countries (2012) \cite{WBSP15} and error behavior.}\label{fig:dual_gdp_oecd_members_2012}\end{figure}
\begin{figure}[!htbp]\flushright\csvtofigrot{newcharts/countries_internet_users_top40_2009.csv}{Classes}{1.2}{\tiny}\includegraphics[width=0.93\linewidth]{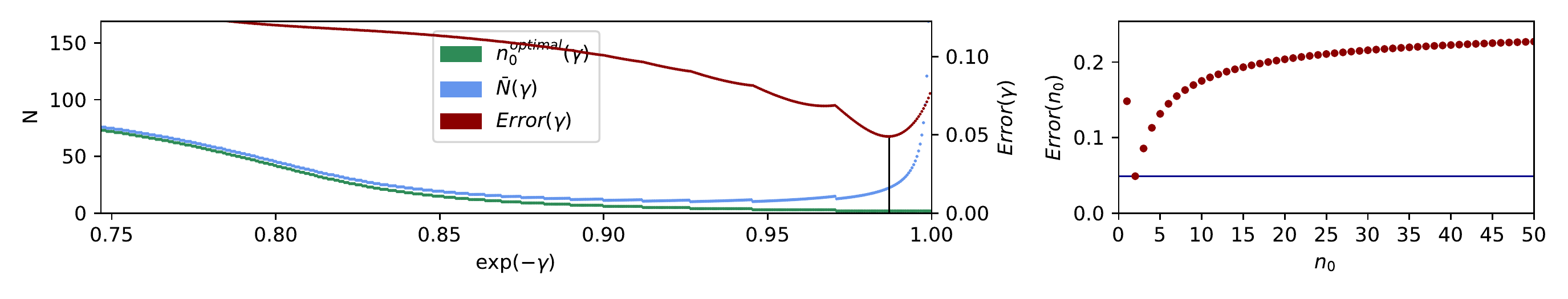}\caption{Number of internet users by country (2009, top 40) \cite{cia-worldfact-14} and error behavior.}\label{fig:dual_countries_internet_users_top40_2009}\end{figure}
\begin{figure}[!htbp]\flushright\csvtofigrot{newcharts/bugs_glibc_2014.csv}{Classes}{2.2}{\tiny}\includegraphics[width=0.93\linewidth]{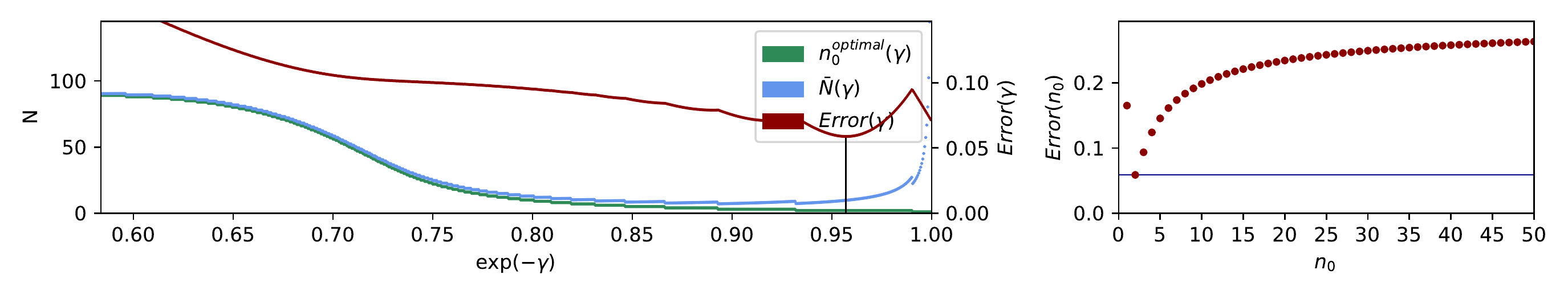}\caption{Number of bugs in glibc, by component (2014) \cite{WBSP6} and error behavior.}\label{fig:dual_bugs_glibc_2014}\end{figure}
\begin{figure}[!htbp]\flushright\csvtofigrot{newcharts/countries_population_top_40_july_2014.csv}{Classes}{1.2}{\tiny}\includegraphics[width=0.93\linewidth]{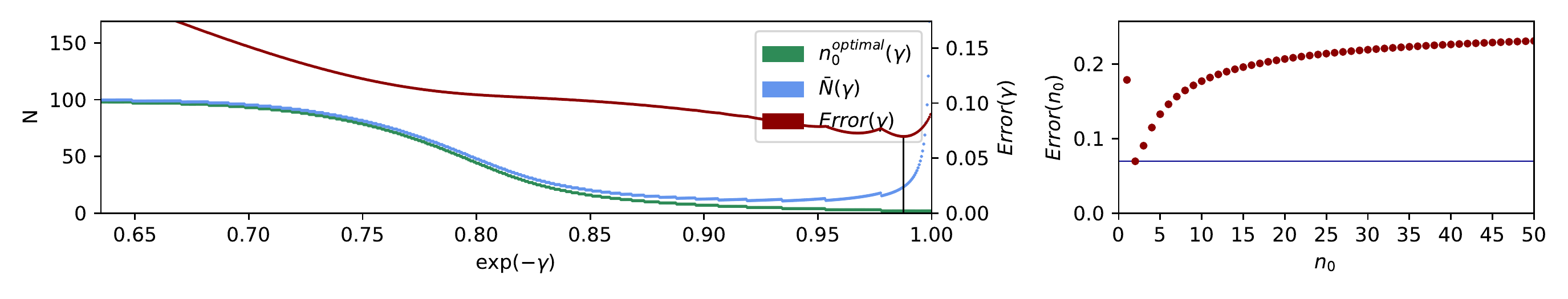}\caption{40 most populated countries in July 2014 \cite{cia-worldfact-14} and error behavior.}\label{fig:dual_countries_population_top_40_july_2014}\end{figure}
\begin{figure}[!htbp]\flushright\csvtofigrot{newcharts/german_motor_vehicles_production_country_2014.csv}{Classes}{3.5}{\tiny}\includegraphics[width=0.93\linewidth]{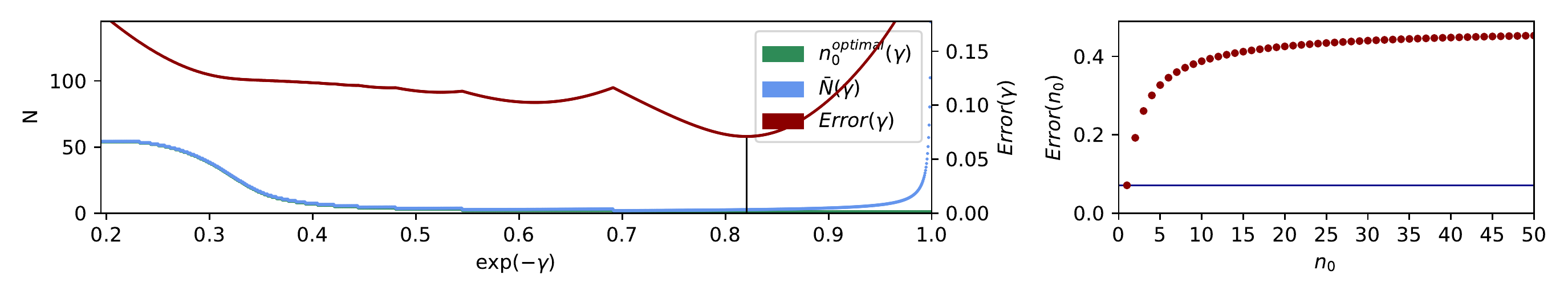}\caption{Registered motor vehicles in Germany by production country (1/1/2014) \cite{WBSP18} and error behavior.}\label{fig:dual_german_motor_vehicles_production_country_2014}\end{figure}
\begin{figure}[!htbp]\flushright\csvtofig{newcharts/euro_cash_2014.csv}{Classes}{4}{\tiny}\includegraphics[width=0.93\linewidth]{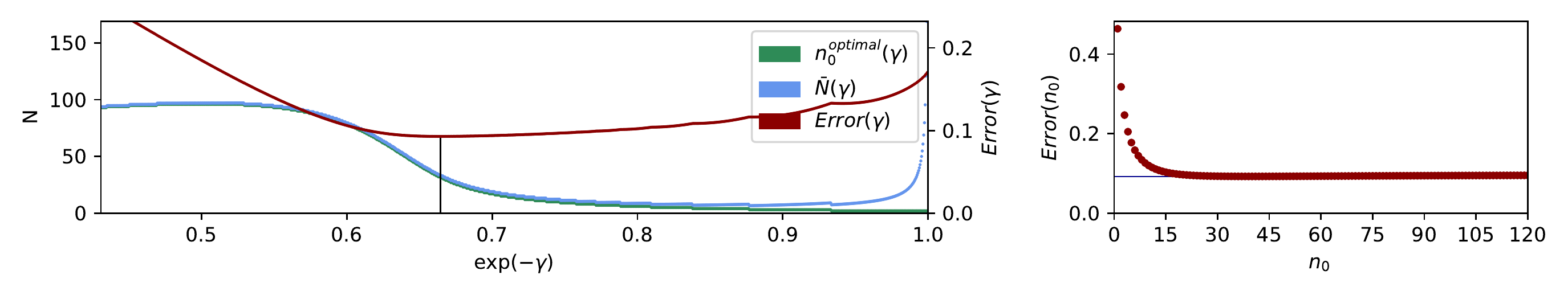}\caption{Euro cash in circulation by coin/note values (2014) \cite{WBSP19} and error behavior.}\label{fig:dual_euro_cash_2014}\end{figure}
\begin{figure}[!htbp]\flushright\csvtofig{newcharts/counterfeit_euro_notes_2013.csv}{Classes}{6}{\footnotesize}\includegraphics[width=0.93\linewidth]{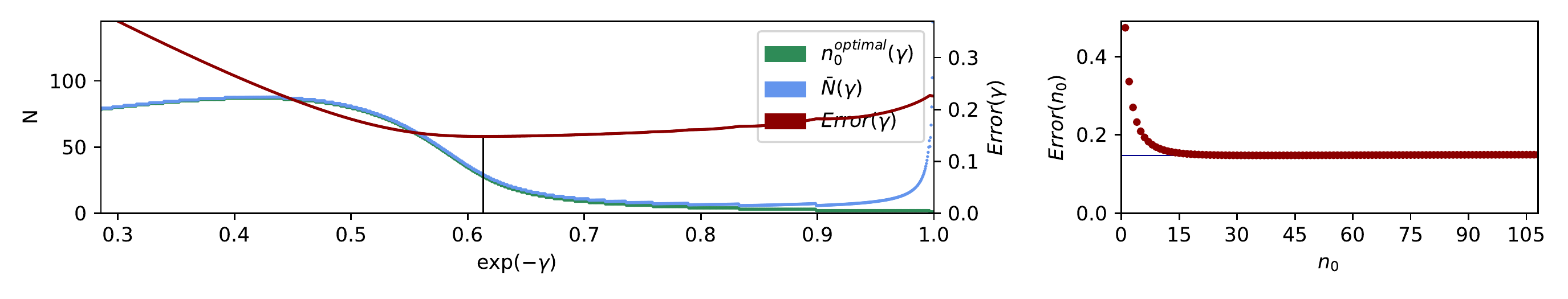}\caption{Counterfeit \euro{}-bills by note value (2013) \cite{WBSP11} and error behavior.}\label{fig:dual_counterfeit_euro_notes_2013}\end{figure}
\begin{figure}[!htbp]\flushright\csvtofigrot{newcharts/wikipedia_popular_pages_top_40_jan_2015.csv}{Classes}{1.2}{\tiny}\includegraphics[width=0.93\linewidth]{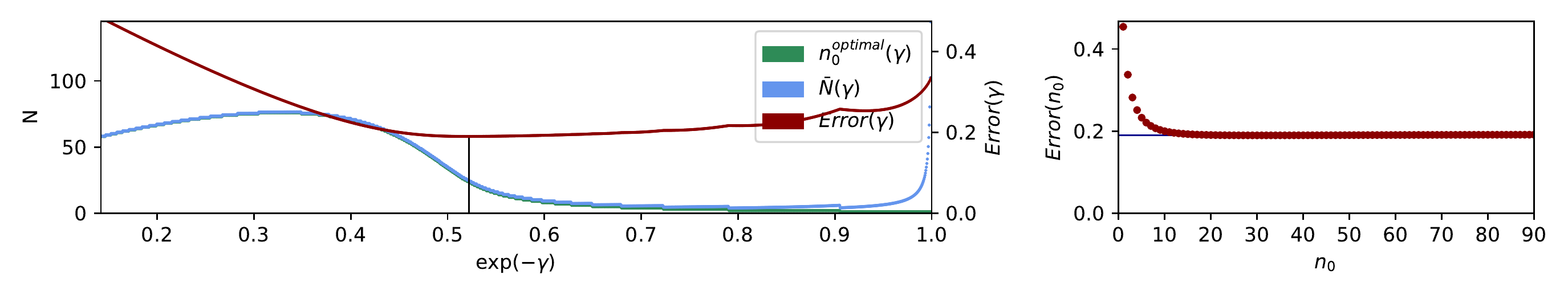}\caption{Top 40 requested pages on wikipedia (January 2015) \cite{WBSP29} and error behavior.}\label{fig:dual_wikipedia_popular_pages_top_40_jan_2015}\end{figure}

\bibliographystyle{unsrt}
\bibliography{literature}


\end{document}